\documentclass[a4paper,11pt]{article}
\pdfoutput=1 

\usepackage{jheppub} 

\usepackage{amsmath}
\usepackage{amssymb}
\usepackage{young}
\usepackage[vcentermath]{youngtab}
\usepackage{dsfont}
\usepackage{braket}
\usepackage{comment}
\usepackage{textgreek} 
\usepackage{simplewick}

\allowdisplaybreaks
\usepackage{mathrsfs}
\usepackage[mathscr]{euscript}
\usepackage[dvipsnames]{xcolor}
\usepackage{tikz}
\usepackage{empheq}
\usepackage[most]{tcolorbox}

\newtcolorbox{empheqboxed}{colback=gray!30, 
 colframe=white,
 width=\textwidth,
 sharpish corners,
 top=-2mm, 
 bottom=0pt
}

\newcommand*{\Scale}[2][4]{\scalebox{#1}{$#2$}}%

\numberwithin{equation}{section}

\def\be{\begin{equation}}
\def\ee{\end{equation}}
\newcommand{\G}{\mathrm{G}}
\newcommand{\lgamma}{\text{\textgamma}}

\newcommand{\hypgeo}[2]{%
  {\vphantom{F}}_{#1}\kern-\scriptspace F_{#2}%
  }

\renewcommand\Im{\mathop{\text{Im}}}

\renewcommand\Re{\mathop{\text{Re}}}

\title{
String correlators on $\boldsymbol{\text{AdS}_3}$: Three-point functions
}

\author[a]{Andrea Dei,}
\author[b]{Lorenz Eberhardt}
\affiliation[a]{Jefferson Physical Laboratory, Harvard University, \\
\hspace*{0.3cm}Cambridge, MA 02138 USA}
\affiliation[b]{School of Natural Sciences, Institute for Advanced Study, \\
\hspace*{0.3cm}Princeton, NJ 08540, USA}
\emailAdd{adei@fas.harvard.edu}
\emailAdd{elorenz@ias.edu}

\abstract{We revisit the computation of string worldsheet correlators on Euclidean $\text{AdS}_3$ with pure NS-NS background. We compute correlation functions with insertions of spectrally flowed operators. We explicitly solve all the known constraints of the model and for the first time conjecture a closed formula for three-point functions with arbitrary amount of spectral flow. We explain the relation of our results with previous computations in the literature and derive the fusion rules of the model. This paper is the first in a series with several installments.}

\begin{document}
\maketitle
\flushbottom

\section{Introduction}
The description of string theory via worldsheet CFT forms the backbone of the theory and in many cases, it is essentially the only way to access dynamical properties such as scattering amplitudes. Even though one can only study string theory perturbatively in this way, this already contains very non-trivial information and string scattering amplitudes quickly become too complicated to compute.

The last couple of years have seen a revival of worldsheet methods in string theory, both in flat space and in curved space. These worldsheet computations have in particular been applied to the holographic setup. Amenable non-trivial backgrounds for string worldsheet theory include in particular the minimal string, the $c=1$ string and string theory on $\text{AdS}_3$ with pure NS-NS flux and our understanding of holography in each of these instances was greatly improved. The $(2,p)$-minimal string is dual to a particular matrix model and the correspondence has been understood by now to an impressive level of detail \cite{Douglas:1989ve, Gross:1989vs, Brezin:1990rb, Saad:2019lba, Mertens:2020hbs}. The $c=1$ string is dual to matrix quantum mechanics and in this context even non-perturbative quantities were successfully matched on both sides of the duality \cite{Klebanov:1991qa, Ginsparg:1993is, McGreevy:2003kb, Douglas:2003up, Balthazar:2019rnh, Balthazar:2019ypi}.\footnote{In this case, the computation of some quantities required string field theory \cite{Sen:2019qqg, Sen:2020oqr, Sen:2020eck}.} Finally, pure NS-NS $\text{AdS}_3$ backgrounds are conjectured to be related to symmetric product orbifold CFTs \cite{Maldacena:1997re}. This connection is sharpest for the superstring on $\text{AdS}_3 \times \text{S}^3 \times \mathbb{T}^4$ with one unit of NS-NS flux where the dual CFT is conjectured to be the symmetric orbifold of the $\mathbb{T}^4$ sigma-model \cite{Gaberdiel:2018rqv, Giribet:2018ada, Eberhardt:2018ouy, Eberhardt:2019ywk, Eberhardt:2020akk, Dei:2020zui, Knighton:2020kuh, Gaberdiel:2020ycd}. In this instance, string perturbation theory truncates and there do not seem to be any non-perturbative phenomena. Thus, string perturbation theory can be used to prove the correspondence perturbatively in this case, a program that is close to completion. Moreover, recent advances suggest that the study of the tensionless $\text{AdS}_3$ string might provide new insights into the $\text{AdS}_5/\text{CFT}_4$ correspondence as well \cite{Gaberdiel:2021iil,Gaberdiel:2021jrv}. 

String theory on pure NS-NS backgrounds can be described on the worldsheet using an $\mathrm{SL}(2,\mathds{R})$ WZW-model. This has a long history and was one of the first curved backgrounds to be investigated. In particular, spacetime unitarity and worldsheet modular invariance were initially not understood until the need to include spectrally flowed sectors into the theory was realised \cite{Balog:1988jb, Petropoulos:1989fc, Hwang:1990aq, Henningson:1991jc, Gawedzki:1991yu, Bars:1995mf, Teschner:1997ft, Evans:1998qu, Giveon:1998ns, deBoer:1998gyt, Kutasov:1999xu, Teschner:1999ug, Giribet:1999ft, Giribet:2000fy, Maldacena:2000hw, Maldacena:2000kv, Maldacena:2001km, Giribet:2001ft}. In string theory such spectrally flowed vertex operators correspond to winding strings that touch the boundary of $\mathrm{AdS}_3$ at the boundary insertion point and can wind several times around it. Thus the main objects of interest are string correlators whose external states are spectrally flowed. In terms of the CFT on the string worldsheet, spectrally flowed vertex operators are described by representations of the current algebra $\mathfrak{sl}(2,\mathds{R})_k$ that are \emph{not} highest weight. This makes it rather hard to compute string correlators with winding states and the problem has not yet been solved satisfactorily.  

An additional motivation to revisit the $\mathrm{SL}(2,\mathds{R})$ WZW-model (or rather its Euclidean analytic continuation that we define more precisely below) comes from the worldsheet CFT itself. In many ways, this CFT is a closely related cousin of Liouville theory, but the complexity is much higher. It exists for a continuum of values that are labelled by a real number $k$, corresponding to the amount of NS-NS flux in the background. $k$ is not quantized because the target space is non-compact. Contrary to compact WZW models, it is an irrational CFT. However the representations are still manageable and one has a chance to completely solve the CFT in the sense of finding an analogue of the DOZZ-formula \cite{Dorn:1994xn, Zamolodchikov:1995aa} for all three-point functions. While this has been achieved for the unflowed sector in \cite{Teschner:1997ft} (that forms a closed subsector of the theory) a general formula that takes into account spectral flow is still missing.\footnote{This is at least true for generic vertex operators that include all the quantum numbers relevant for holography. There are closed formulas for some special correlation functions \cite{Fateev,Maldacena:2000kv,Giribet:2000fy,Giribet:2001ft,Giribet:2005ix,Ribault:2005ms,Giribet:2005mc,Minces:2005nb,Iguri:2007af,Baron:2008qf,Iguri:2009cf,Giribet:2011xf,Cagnacci:2013ufa,Giribet:2015oiy,Giribet:2019new,Hikida:2020kil}.} Finally, there is also a direct relation between correlators of the $H_3^+$ model (that is related to the unflowed sector of the worldsheet theory) and Liouville correlators (this is a different relation than the one of \cite{Eberhardt:2019qcl}, which relates string theory to a \emph{spacetime} Liouville theory), the so-called $H_3^+$-Liouville correspondence \cite{Teschner:1997ft, Stoyanovsky:2000pg, Ribault:2005wp, Giribet:2005ix, Ribault:2005ms, Hikida:2007tq}. This correspondence allows one in particular to express the unflowed $H_3^+$ correlators
in terms of Liouville correlators. We are not aware of a generalization of this correspondence to spectrally flowed correlators.\footnote{The generalization considered in \cite{Giribet:2005ix,Ribault:2005ms,Giribet:2015oiy,Giribet:2019new} only deals with correlators for which spectral flow simplifies dramatically. See Section \ref{sec:spectral flow violation}. }

Let us clarify what we mean exactly by the worldsheet model. There has been some naming confusion in the literature. The analytic continuation of the $\mathrm{SL}(2,\mathds{R})_k$ WZW describes the model one obtains from Wick rotating in field space the $\mathrm{SL}(2,\mathds{R})_k$ WZW model. Actually, the theory one wants has the universal covering space of $\widetilde{\mathrm{SL}}(2,\mathbb{R})_k$ as its target space. We will follow the by now standard convention to omit the tilde. We always mean the WZW model on the universal covering in the following. The analytic continuation is not the sigma model with Euclidean $\mathrm{AdS}_3$ as target space, since normalizability in the sigma-model is still defined with respect to the Lorentzian target space. As such the spectrum of the original $\mathrm{SL}(2,\mathds{R})_k$ coincides with the spectrum of the analytic continuation. The CFT with Euclidean target space is usually called $H_3^+$ model and involves only continuous unflowed representations. Teschner showed that the $H_3^+$ model is a crossing-symmetric CFT \cite{Teschner:1997ft, Teschner:1999ug, Teschner:2001gi}. We are inspired by string theory on Euclidean $\text{AdS}_3$, which is defined by analytic continuation from Lorentzian signature and in particular contains spectrally flowed representations. We understand in the following the worldsheet model as the analytic continuation of the (universal cover of the) $\mathrm{SL}(2,\mathds{R})_k$ WZW model. 

\medskip

We are thus motivated to revisit the worldsheet model (including spectrally flowed sectors) to address these problems. Since we want to apply the gained knowledge to string theory, we consider vertex operators in the so-called $x$-basis. In this basis, worldsheet vertex operators depend besides the worldsheet coordinate $z$ on a second coordinate $x$ that corresponds to the insertion point of the string on the boundary of Euclidean $\mathrm{AdS}_3$ (that is given by a Riemann sphere). String theory on $\text{AdS}_3$ is conjectured to be dual to a 2D CFT living on the boundary sphere \cite{Maldacena:1997re} --- see e.g.~\cite{David:2002wn} for a review. From the point of view of the dual CFT, $x$ labels the position coordinate where vertex operators are inserted. The vertex operators of the worldsheet theory depend additionally on an integer $w \ge 0$ that is the spectral flow. They only transform in a highest weight representation of the current algebra $\mathfrak{sl}(2,\mathds{R})_k$ for $w=0$. For more details on how they are defined, we refer to Section~\ref{sec:spectral flow}. Moreover, vertex operators depend on an $\mathfrak{sl}(2,\mathds{R})_k$ spin $j$ that can take values either in the discrete series $j \in \mathds{R}$ or in the principal continuous series of $j \in \frac{1}{2}+i \mathds{R}$ of $\mathfrak{sl}(2,\mathds{R})$. Finally, vertex operators transform under global $\mathrm{SL}(2,\mathds{R})$ transformations with weight $(h,\bar{h})$, which is identified with the conformal dimension of the dual CFT. To summarize, the correlators we want to study take the form 
\be 
\left\langle \prod_{i=1}^n V^{w_i}_{j_i, h_i,\bar{h}_i}(x_i; z_i)\right\rangle \ . 
\label{intro-flowed-correlator}
\ee
This is to be compared to correlators of unflowed vertex operators, i.e.~correlators of the form 
\be 
\left\langle \prod_{i=1}^n V^{0}_{j_i}(x_i; z_i)\right\rangle \ , 
\label{intro-unflowed-correlator}
\ee
which have been investigated thoroughly in the literature \cite{Teschner:1997ft, Teschner:1999ug, Teschner:2001gi}.  

We want to highlight a further subtlety that also has caused confusion. The choice of vertex operators that results from the $x$-basis corresponds to coherent states in the Verma module of $\mathfrak{sl}(2,\mathds{R})_k$. Such states are usually not considered in CFT, where one mostly only considers finite linear combinations of basis elements in the Verma module. It turns out that very important aspects of the model depend on this choice. In particular the fusion rules that one obtains in this way are less restrictive than for the more conservative choice where only finite linear combinations are considered. We are interested in string theory where such coherent states are clearly necessary to describe the correct physics. Therefore there are more three-point functions to be determined to give a complete solution of the model. We can always recover results that have been obtained without allowing for coherent states by taking suitable limits that correspond to the collision of vertex operators in the $x$-space. We explain this relation very carefully in this paper.

The unflowed vertex operators entering \eqref{intro-unflowed-correlator} carry no explicit dependence on $h_i$. This is because for $w=0$ the spin $j$ and the spacetime conformal weight $h$ are identified, $h=\bar{h}=j$. Already at this stage, one of the difficulties hampering the study of correlators with spectral flow insertions becomes manifest: correlators of the form \eqref{intro-flowed-correlator} depend on $n$ additional parameters with respect to \eqref{intro-unflowed-correlator}. As we mentioned previously, an additional obstacle comes from the fact that flowed vertex operators are not highest weight representations of the current algebra on the worldsheet. Since positive modes don't annihilate the vertex operators, Ward identities contain unknown terms of the form, 
\be 
\left\langle J^+_p V_{j_i,h_i,\bar{h}_i}^{w_i}(x_i;z_i) \prod_{\ell \neq i} V_{j_\ell,h_\ell,\bar{h}_\ell}^{w_\ell}(x_\ell;z_\ell) \right\rangle \ , \qquad 0 < p < w_i \ .  
\label{intro-unknowns}
\ee
This issue has been solved in \cite{Eberhardt:2019ywk}, where the authors showed that one can consistently eliminate the unknowns \eqref{intro-unknowns} from the Ward identities and deduce $n$ additional constraints for the correlators \eqref{intro-flowed-correlator}. These constraints are usually involved: they amount to a system of coupled recursion relations in the spacetime CFT conformal weights $h_i$, involving derivatives with respect to the boundary coordinates $x_i$. An explicit solution of these constraints is known only for special combinations of the spins $j_i$, see \cite{Eberhardt:2019ywk} for more details. 

\medskip 

The core of the paper is devoted to the study of the recursion relations of \cite{Eberhardt:2019ywk} for three-point functions. For generic choice of the spins $j_i$, we explicitly solve these constraints  in many examples and propose a closed formula for three-point functions with insertions of operators carrying an arbitrary amount of spectral flow. One of our main innovations is to introduce a new basis for the vertex operators in terms of which the expressions simplify drastically. This basis is obtained by an integral transform in the variables $(h,\bar{h})$. Our main result in this so-called $y$-basis is given in eq.~\eqref{eq:main conjecture} and takes a surprisingly simple form.

The solution that we find is non-trivial and can be written in terms of generalized hypergeometric functions when transformed back to the original basis in terms of conformal weights $(h,\bar{h})$. It looks much more complicated in the $h$-basis than in the $y$-basis. For example, when all vertex operators sit in lowest weight discrete representations (these representations will be explained in Section~\ref{sec:bosonic strings}) with odd $\sum_{i=1}^3 w_i$ and $\sum_{i=1}^3 w_i >2 \max_{i} w_i$, the solution can be written in terms of the so-called Lauricella hypergeometric function,\footnote{We find it intriguing that the same Lauricella hypergeometric function shows up in the Feynman rules for scalar conformal correlators in Mellin space \cite{Fitzpatrick:2011ia, Paulos:2011ie}. }
\begin{align}
&\left\langle V_{j_1,h_1,\bar{h}_1}^{w_1}(0;0)V_{j_2,h_2,\bar{h}_2}^{w_2}(1;1)V_{j_3,h_3,\bar{h_3}}^{w_3}(\infty;\infty)\right\rangle =\mathcal N(j_1) D(\tfrac{k}{2}-j_1,j_2,j_3)  \nonumber\\
&\qquad\times(N')^{k-2j_1-2j_2-2j_3}\prod_{i=1}^3 a_i^{\frac{k}{2}(w_i-1)-h_i-\bar{h}_i}w_i^{-\frac{k}{2}(w_i+1)+2j_i} \Pi^{-k}  \nonumber\\
&\qquad\times \left|\prod_{i=1}^3 \binom{j_i-\frac{kw_i}{2}+h_i-1}{ 2j_i-1} F_A\left(j_1+j_2+j_3-\frac{k}{2};j_i-h_i+\frac{k w_i}{2};2j_i;\frac{w_i}{N'} \right)\right|^2\ .
 \label{eq:intro three point function}
\end{align}
The constant prefactor $\mathcal N(j_1)D(\tfrac{k}{2}-j_1,j_2,j_3)$ is determined conjecturally. Here, $\mathcal N(j)$ is a normalization factor of the vertex operators that, up to factors depending only on $k$, appears already in the unflowed two-point function \eqref{eq:normalization vertex operators}. $D(j_1,j_2,j_3)$ is the unflowed three-point function \eqref{eq:unflowed three point function}. Furthermore, $N'=\frac{1}{2}(w_1+w_2+w_3+1)$ and $a_i$ is a simple ratio of binomial coefficients given in \eqref{eq:three-pt-ai} while $\Pi$ is given in \eqref{eq:definition C}. We should also mention that the form of the correlator changes drastically depending on $\sum_i w_i$ being even or odd.
We obtain similar results for other representations, see eqs.~\eqref{h-basis-solution-3D-} and \eqref{h-basis-3pt-C}.
We also (re)derive the fusion rules of the model. They depend crucially on the fact that we are considering vertex operators that depend on an additional $x$-coordinate. 

\medskip
This paper is the first in a series. In a future publication \cite{paper2} we will generalize our findings to four-point functions and discuss possible implications for the spacetime CFT. In this forthcoming paper we will interpret our results much more geometrically than here.

\medskip

The paper is organized as follows. We start in Section~\ref{sec:bosonic strings} to review and develop some necessary background on the description of string theory on $\mathrm{AdS}_3$ with pure NS-NS flux, the $\mathrm{SL}(2,\mathds{R})_k$ model and its analytic continuation. We then review global and local Ward identities in Section~\ref{sec:constraints}. We start by applying these constraints to two-point functions in Section~\ref{sec:2ptf} and see that they are completely fixed up to overall normalization. We then move on to three point functions in Section~\ref{sec:3ptf}. While their dependence on the worldsheet and spacetime coordinates is still trivially fixed by the global Ward identities, their dependence on the spacetime conformal weights is very non-trivial, for which we give a closed form expression in terms of a hypergeometric integral. In special cases one can evaluate this hypergeometric integral explicitly in terms of the Lauricella hypergeometric function of type A. As a byproduct of our analysis, we give a new derivation of the fusion rules of the model. We make some examples and comment on the relation of our work with previous results in the literature. We conclude in Section \ref{sec:conclusions}, where we discuss some open questions and propose new directions of research. We attach to this paper an ancillary {\tt Mathematica} notebook. The reader can use it to reproduce the computations that are discussed in the following sections.  

\section{Bosonic strings on Euclidean \texorpdfstring{$\text{AdS}_{\boldsymbol 3}$}{AdS3}}
\label{sec:bosonic strings}

In this paper we study correlators of bosonic strings on Euclidean $\text{AdS}_3$.  The worldsheet theory is defined by analytic continuation of the $\mathrm{SL}(2,\mathds{R})_k$ WZW model in a coherent state basis as mentioned in the Introduction. Following \cite{Gawedzki:1991yu,Teschner:1997ft,Giveon:1998ns,deBoer:1998gyt,Kutasov:1999xu,Teschner:1999ug,Maldacena:2000hw, Maldacena:2000kv, Maldacena:2001km}, we will describe the worldsheet theory in terms of the analytically continued $\text{SL}(2,\mathds{R})_k$ WZW model. Let us review its main features, which will be of primary interest in our study. 

\subsection[Spectral flow of \texorpdfstring{$\mathfrak{sl}(2,\mathds{R})_k$}{sl(2,R)k}]{Spectral flow of $\boldsymbol{\mathfrak{sl}(2,\mathds{R})_k}$}
\label{sec:spectral flow}

The $\mathfrak{sl}(2,\mathds R)_k$ algebra is defined by the generators $J_n^a$ with $a \in \{\pm, 3\}$, which satisfy the commutation relations
\be
[J^3_m, J^3_n] = -\tfrac{1}{2} \, k \, m \, \delta_{m+n,0} \ , \quad
[J^3_m, J^\pm_n]  = \pm J^\pm_{m+n} \ , \quad
[J^+_m, J^-_n] = k \, m \, \delta_{m+n,0} - 2 J^3_{m+n} \ . 
\ee
For any non-negative integer $w$, this algebra admits a spectral flow automorphism $\sigma^w$, which acts on the modes as
\be 
\sigma^w(J^\pm_m) = J^\pm_{m\mp w} \ , \qquad \sigma^w(J^3_m) = J^3_m + \frac{kw}{2}\, \delta_{m,0} \ . 
\label{spectral-flow-automorphism}
\ee
Highest weight representations of the affine algebra are obtained as usual by acting with negative modes on $\mathfrak{sl}(2,\mathds R)$ modules of the zero modes. 
Two families of $\mathfrak{sl}(2,\mathds{R})$ representations will be of interest for us:
\begin{itemize}
\item \emph{Discrete representations}, characterized by $j \in \mathds R$ and  $m-j \in \mathds N_0$
\begin{subequations}
\begin{align}
& \text{Lowest weight} & & \mathcal{D}_j^+ = \{\ket{j,m} :  m = j, j+1, j+2, \dots  \} 
\label{DiscreteD+} \\
& \text{Highest weight} & & \mathcal{D}_j^- = \{\ket{j,m} :  m = -j, -j-1, -j-2, \dots  \} \label{DiscreteD-} 
\end{align} 
\end{subequations}
\item \emph{Continuous representations} $\mathcal{C}_j$, for which $j \in \tfrac{1}{2} + i \, \mathds R$ and $m \in \mathds Z + \lambda$.\footnote{Usually the continuous representations are labelled by the two parameters $j$ and $\lambda$. We will omit $\lambda$ in the following, because it is fully specified by the quantum number $h$ that we introduce below. Thus we only write $\mathcal{C}_j$.}
\end{itemize} 
For both discrete and continuous representations, the Casimir is given by
\be 
\text{Cas} = - j(j-1) \ . 
\label{casimir}
\ee
We sometimes abuse notation and denote also the full affine representation that is built on top of these global representations by $\mathcal{D}_j^\pm$ and $\mathcal{C}_j$.
Additional affine representations can be obtained by composing the action of the $J^a_n$ modes with the spectral flow automorphism \eqref{spectral-flow-automorphism}. They can also be defined in terms of `affine primary' states $[\ket{j,m}]^w$ as follows,\footnote{Compared to \cite{Eberhardt:2019ywk}, we redefined $j \to 1-j$, which makes our formulas uniform for the unflowed and the flowed sector. See also the discussion at the end of Section \ref{sec:local-ward-id}. }  
\begin{subequations}
\begin{align}
J_w^+ \bigl[\ket{j,m}\bigr]^w &= (m+1-j) \bigl[\ket{j,m+1}\bigr]^w \ , & J^+_n \bigl[\ket{j,m}\bigr]^w & = 0 \ , \quad  n> w \ , \\
J_0^3 \bigl[\ket{j,m}\bigr]^w &= \left(m+\tfrac{k\, w}{2}\right) \bigl[\ket{j,m}\bigr]^w \ , & J^3_n \bigl[\ket{j,m}\bigr]^w & = 0 \ , \quad  n> 0 \ , \\
J_{-w}^- \bigl[\ket{j,m}\bigr]^w &= (m-1+j) \bigl[\ket{j,m-1}\bigr]^w \ , & J^-_n \bigl[\ket{j,m}\bigr]^w & = 0 \ , \quad  n> -w \ .
\end{align}
\label{flowed-rep}
\end{subequations}
We will refer to these representations as \emph{spectrally flowed representations}. These actions can be consistently truncated to the ranges of the discrete representations. We denote the resulting representations by $[\mathcal{C}_j]^w$ and $[\mathcal{D}_j^\pm]^w$. There is furthermore the following identification on affine spectrally flowed representations \cite{Maldacena:2000hw}:
\be 
[\mathcal{D}_j^+]^w=[\mathcal{D}_{\frac{k}{2}-j}^-]^{w+1}\ , \label{eq:identification representations D+ D-}
\ee
that we shall use repeatedly. This can be read off from Figure~\ref{fig:Verma module} below.

\paragraph{Spectrum of the $\boldsymbol{\mathrm{SL}(2,\mathds{R})_k}$ WZW model.} After having established the relevant representations, we need to decide which representations appear in the Hilbert space of the $\mathrm{SL}(2,\mathds{R})_k$ WZW model. We want to analyze the same representations in the analytic continuation to Euclidean spacetime signature. This was established by Maldacena and Ooguri \cite{Maldacena:2000hw} and their proposal is to include all continuous representations $[\mathcal{C}_j]^w$ with $w \ge 0$ and $j \in \frac{1}{2}+i \mathds{R}$ and all discrete representations $[\mathcal{D}_j^+]^w$ with $w \ge 0$ and $\frac{1}{2}<j<\frac{k-1}{2}$. Due to the identification \eqref{eq:identification representations D+ D-}, this automatically also includes all representations $[\mathcal{D}_j^-]^w$ with $w \ge 1$.\footnote{Often string theory is treated in a basis where $J_0^3$ is diagonalized. In this case, one also has to include negative spectral flow. For the vertex operators that we will introduce momentarily, this is not the case because the transformation $x \to \frac{1}{x}$ will flip the spectral flow $w \to -w$. Thus, we restrict to non-negative spectral flow.} For these representations a no-ghost theorem can be proven \cite{Hwang:1990aq, Evans:1998qu, Maldacena:2000hw}.

\subsection[Vertex operators in the \texorpdfstring{$x$}{x}-basis]{Vertex operators in the $\boldsymbol{x}$-basis}
\label{x-basis}

While the results of this paper will mainly concern bulk correlators, our interest will be geared towards holographic applications. When defining vertex operators of primary states it is then natural to introduce the dependence on the dual spacetime coordinate $x$. Of course, vertex operators also depend on the worldsheet coordinate $z$ and we have 
\be 
V_{j,h, \bar h}^w(x;z) \equiv e^{z L_{-1}} \, e^{x J_0^+} \, V_{j,h, \bar h}^w(0;0) \, e^{-x J_0^+} \, e^{-z L_{-1}} \ , 
\label{x-basis-vertex-op}
\ee  
where we have made use of the fact that $L_{-1}$ and $J_0^+$ are respectively the translation operator on the worldsheet and in spacetime. Notice that since $L_{-1}$ and $J_0^+$ commute, $[L_{-1},J_0^+]=0$, there is no ordering ambiguity in \eqref{x-basis-vertex-op}. Obviously, the vertex operator \eqref{x-basis-vertex-op} also depends on the anti-holomorphic coordinates $\bar x$ and $\bar z$. This is implemented by the action of the anti-holomorphic translation operators $\bar J^+_0$ and $\bar L_{-1}$, respectively on the worldsheet and in spacetime. Here and in the following we will frequently omit the explicit anti-holomorphic dependence of vertex operators. The vertex operator $V^w_{j,h,\bar h}(x;z)$ is associated to the state 
\be 
[\ket{j,m} \otimes \ket{j, \bar m}]^{w} = V_{j,h, \bar h}^w(0;0)\ket{0}  \ ,  
\ee
where the spacetime conformal dimension $h$ is given by the $J_0^3$ eigenvalue, 
\be 
J_0^3 \, [\ket{j,m}]^{w} = h \, [\ket{j,m}]^{w} = \Bigl(m + \frac{kw}{2} \Bigr) \, [\ket{j,m}]^{w} \ . 
\ee
The identification $h=m+\frac{kw}{2}$ follows from eq.~\eqref{spectral-flow-automorphism}. 
Similar formulas apply for the anti-holomorphic spacetime conformal dimension $\bar h$. In order to lighten the notation, we will frequently write $V^w_{j,h}(x;z)$ instead of $V^w_{j, h , \bar h}(x;z)$.
The defining OPEs of the $\mathfrak{sl}(2,\mathds{R})$ currents with the $V_{j,h}^w(x;z)$ read
\begin{subequations} \label{eq:x-OPEs}
\begin{align}
J^+(\zeta) V_{j,h}^w(x;z)&=\sum_{p=1}^{w+1} \frac{(J^+_{p-1} V_{j,h}^w)(x;z)}{(\zeta-z)^p}+\mathcal{O}(1)\ , \\
\big(J^3(\zeta)-x J^+(\zeta)\big) V_{j,h}^w(x;z)&=\frac{h V_{j,h}^w(x;z)}{\zeta-z}+\mathcal{O}(1)\ , \\
\big(J^-(\zeta)-2x J^3(\zeta)+x^2 J^+(\zeta)\big)  V_{j,h}^w(x;z)&=(\zeta-z)^{w-1} (J^-_{-w} V_{j,h}^w)(x;z)+\mathcal{O}((\zeta-z)^{w})\ .
\label{regular-combination-OPE}
\end{align}
\end{subequations}
The shifts on the left-hand-side of these OPEs can be derived from \eqref{x-basis-vertex-op}. From the previous discussion we also have the identifications
\begin{subequations}
\begin{align}
(J_0^+ V_{j,h}^w)(x;z)&=\partial_x V_{j,h}(x;z)\ , \label{eq:J0p action}\\
(J_{\pm w}^+ V_{j,h}^w)(x;z)&=\left(h-\tfrac{k}{2}w\pm (1-j)\right) V_{j,h\pm 1}^w(x;z)\ .\label{eq:Jwp action}
\end{align}
\end{subequations}
In the unflowed sector, these OPEs reduce to
\begin{subequations}
\begin{align}
J^+(\zeta) V_{j}^0(x;z) &= \frac{\partial_x V_{j}^0(x; z)}{\zeta-z} +\mathcal{O}(1) \ , \\
J^3(\zeta) V_{j}^0(x;z) &= \frac{(j+x\partial_x) V_{j}^0(x; z)}{\zeta-z}  +\mathcal{O}(1) \ , \\
J^-(\zeta) V_{j}^0(x;z) &= \frac{(2  j x + x^2\partial_x) V_{j}^0(x; z)}{\zeta-z} +\mathcal{O}(1) \ . 
\end{align}
\label{eq:x-OPEs unflowed}
\end{subequations}
The OPEs \eqref{eq:x-OPEs}  and \eqref{eq:x-OPEs unflowed} define the vertex operators and essentially everything we say in this paper is a consequence of these fundamental OPEs.

It is in principle straightforward to also consider vertex operators of descendant fields, although we will not do so in this paper. See \cite{Bertle:2020sgd} for some recent result in this direction. 

\subsection{Reflection symmetry}

Unflowed continuous representations with spin $j$ and $1-j$ are equivalent and the associated $x$-basis vertex operators are identified \cite{Teschner:1997ft},
\be 
V^0_{1-j}(x;z) = R_{1-j} \frac{(1-2j)}{\pi} \int \mathrm{d}^2 x' \ |x-x'|^{4j-4} \,  V^0_{j}(x';z) \ . 
\label{eq:Teschner-reflection}
\ee
Here, $R_j$ is the unflowed reflection coefficient.
For flowed continuous representation, the vertex operators $V^w_{j,h}(x;z)$ and $V_{1-j,h}^w(x;z)$ are identified and we write 
\be 
V_{1-j, h, \bar h}^w(x;z) = R(1-j,h,\bar h) \, V_{j,h,\bar h}^w(x;z) \ ,
\label{reflection-symm} 
\ee
where we have introduced the \emph{reflection coefficient} $R(j,h, \bar h)$, allowing for an arbitrary relative normalization of the two vertex operators. We will refer to eq.~\eqref{reflection-symm} as \emph{reflection symmetry}. Idempotency requires 
\be 
R(j,h, \bar h) \, R(1-j,h, \bar h) = 1  \ .
\label{Rj-idempotency} 
\ee
The $h$ and $\bar h$ dependence of the reflection coefficient is constrained by the choice of the representation. In fact, 
\begin{multline}
R(1-j,h,\bar h) \, (h-\tfrac{k \, w}{2}+1-j) \, V_{j, h+1, \bar h}^w(x;z) = R(1-j,h,\bar h) \,  J^+_w V_{j, h, \bar h}^w(x;z)
\\ = J^+_w  V_{1-j, h, \bar h}^w(x;z) 
 = (h-\tfrac{k \, w}{2}+j) \, R(1-j, h+1, \bar h) \, V_{j, h+1, \bar h}^w(x;z)
\end{multline}   
implies a recursion relation for $R(j,h,\bar h)$. It follows
\be 
R(j,h,\bar h) = R(j, \lambda) \,  \binom{h-\tfrac{k \, w}{2}+j-1}{2j-1} \, \binom{\bar h-\tfrac{k \, w}{2}+j-1}{2j-1} \ , 
\label{eq:Rjlambda}
\ee
where $R(j, \lambda)$ is an arbitrary function of $j$ and $\lambda = ( h-\tfrac{kw}{2}) \text{ mod } \mathds{Z} = (\bar h -\tfrac{kw}{2}) \text{ mod } \mathds{Z}$. 
In the following, we will find it useful to also specify the $\lambda$ dependence of the reflection coefficient,
\be 
R(j,\lambda) = \frac{\pi \, R_{j} \, (1-2j)  }{\sin (2 \pi j)} \frac{\sin \left(\pi(\lambda +j) \right)}{\sin \left(\pi(\lambda-j) \right)} \ , 
\label{hmodZ-reflection-coeff}  
\ee
where $R_j$ is an arbitrary function of $j$ satisfying
\be
R_j \, R_{1-j} = 1 \ . 
\label{eq:Rj-R1-j}
\ee 
Notice that this is just a choice of convenience, simplifying many of the formulae appearing in the following sections. 
It is convenient to write 
\be 
R(j,h,\bar{h})=R_{j} (2j-1) \frac{\lgamma(h-\frac{kw}{2}+j)}{\lgamma(h-\frac{kw}{2}+1-j)\lgamma(2j)}\ ,
\label{eq:reflection coeff and gamma}
\ee
where
\be 
\lgamma(x)=\frac{\Gamma(x)}{\Gamma(1-\bar{x})}\ . \label{eq:gamma function}
\ee
In this formula, $\bar{x}$ is not the complex conjugate, but the corresponding quantity describing the right-movers. One can furthermore fix the reflection coefficient $R_j$ to be \cite{Teschner:1997ft}
\be 
R_j=\frac{(k-2)\nu^{1-2j}}{(2j-1)\lgamma\left(\frac{2j-1}{k-2}\right)}\ .
\ee
This is the natural choice that comes from the normalization of the vertex operators using the free-field Wakimoto variables in the unflowed sector. See also appendix~\ref{app:unflowed three-point functions} for a slightly longer discussion on this. $\nu$ is here an undetermined constant that can be thought of as the worldsheet cosmological constant. It is undetermined in the Wakimoto variables and we will leave it unfixed.
\subsection[The \texorpdfstring{$y$}{y}-transform]{The $\boldsymbol y$-transform}
\label{sec:y-transform}

\begin{figure}
\begin{center}
\begin{tikzpicture}
\clip(-3.4,-3.4) rectangle (11.4,4.4);

\draw[thick,->] (9,.5) -- (10,.5) node[right] {$J^3_0$};
\draw[thick,->] (9,.5) -- (9,1.5) node[above] {$L_0$};


\draw[gray,xstep=0.5,ystep=0.5, opacity=0.3] (-1.9,-3.4) grid (1.4,3.9);

\draw[cyan,thick,path fading=west, line width=3 ] (-1.7,2.4) -- (-0.5,0);

\draw[cyan,thick,path fading=east, line width=3 ] (-0.5,0) -- (1.2,-3.4);

\draw[magenta,thick,path fading=east, line width=3 ] (-0.5,0) -- (1.4,0);

\foreach \i in {3.5,3,2.5,2}{
\node[circle, fill,draw,scale=.3] at ({-1.5},{\i}) {};}

\foreach \i in {3.5,3,2.5,2,1.5,1}{
\node[circle, fill,draw,scale=.3] at ({-1},{\i}) {};}

\foreach \i in {3.5,3,2.5,2,1.5,1,0.5,0}{
\node[circle, fill,draw,scale=.3] at ({-0.5},{\i}) {};}

\foreach \i in {3.5,3,2.5,2,1.5,1,0.5,0,-0.5,-1}{
\node[circle, fill,draw,scale=.3] at ({0},{\i}) {};}

\foreach \i in {3.5,3,2.5,2,1.5,1,0.5,0,-0.5,-1,-1.5,-2}{
\node[circle, fill,draw,scale=.3] at ({0.5},{\i}) {};}

\foreach \i in {3.5,3,2.5,2,1.5,1,0.5,0,-0.5,-1,-1.5,-2,-2.5,-3}{
\node[circle, fill,draw,scale=.3] at ({1},{\i}) {};}

\draw[->, magenta, line width=0.75] (-0.45,0.15) -- (-0.05,0.15);
\draw[->, magenta, line width=0.75] (0.05,0.15) -- (0.45,0.15);
\draw[->, magenta, line width=0.75] (0.55,0.15) -- (0.95,0.15);

\node[] at (-0.25,0.35) {$\Scale[0.7]{J^+_0}$};
\node[] at (0.25,0.35) {$\Scale[0.7]{J^+_0}$};
\node[] at (0.75,0.35) {$\Scale[0.7]{J^+_0}$};

\draw[<-, cyan, line width=0.75] (-1.65,1.85) -- (-1.225,1);
\draw[<-, cyan, line width=0.75] (-1.15,0.85) -- (-0.725,0);
\draw[->, cyan, line width=0.75] (-0.65,-0.15) -- (-0.225,-1);
\draw[->, cyan, line width=0.75] (-0.15,-1.15) -- (0.275,-2);
\draw[->, cyan, line width=0.75] (0.35,-2.15) -- (0.775,-3);

\node[] at (-1.7,1.4) {$\Scale[0.7]{J^-_{-2}}$};
\node[] at (-1.2,0.4) {$\Scale[0.7]{J^-_{-2}}$};
\node[] at (-0.7,-0.6) {$\Scale[0.7]{J^+_{2}}$};
\node[] at (-0.2,-1.6) {$\Scale[0.7]{J^+_{2}}$};
\node[] at (0.3,-2.6) {$\Scale[0.7]{J^+_{2}}$};


\draw[gray,xstep=0.5,ystep=0.5, opacity=0.3] (3.6,-3.4) grid (6.9,3.9);

\draw[cyan,thick,path fading=east, line width=3 ] (5,0) -- (6.7,-3.4);

\draw[magenta,thick,path fading=east, line width=3 ] (5,0) -- (6.9,0);

\draw[orange,thick,path fading=west, line width=3 ] (5,0) -- (3.8,3.6);

\foreach \i in {3.5,3}{
\node[circle, fill,draw,scale=.3] at ({4},{\i}) {};}

\foreach \i in {3.5,3,2.5,2,1.5}{
\node[circle, fill,draw,scale=.3] at ({4.5},{\i}) {};}

\foreach \i in {3.5,3,2.5,2,1.5,1,0.5,0}{
\node[circle, fill,draw,scale=.3] at ({5},{\i}) {};}

\foreach \i in {3.5,3,2.5,2,1.5,1,0.5,0,-0.5,-1}{
\node[circle, fill,draw,scale=.3] at ({5.5},{\i}) {};}

\foreach \i in {3.5,3,2.5,2,1.5,1,0.5,0,-0.5,-1,-1.5,-2}{
\node[circle, fill,draw,scale=.3] at ({6},{\i}) {};}

\foreach \i in {3.5,3,2.5,2,1.5,1,0.5,0,-0.5,-1,-1.5,-2,-2.5,-3}{
\node[circle, fill,draw,scale=.3] at ({6.5},{\i}) {};}

\draw[->, cyan, line width=0.75] (4.85,-0.15) -- (5.275,-1);
\draw[->, cyan, line width=0.75] (5.35,-1.15) -- (5.775,-2);
\draw[->, cyan, line width=0.75] (5.85,-2.15) -- (6.275,-3);

\node[] at (4.8,-0.6) {$\Scale[0.7]{J^+_{2}}$};
\node[] at (5.3,-1.6) {$\Scale[0.7]{J^+_{2}}$};
\node[] at (5.8,-2.6) {$\Scale[0.7]{J^+_{2}}$};

\draw[->, magenta, line width=0.75] (5.05,0.15) -- (5.45,0.15);
\draw[->, magenta, line width=0.75] (5.55,0.15) -- (5.95,0.15);
\draw[->, magenta, line width=0.75] (6.05,0.15) -- (6.45,0.15);

\node[] at (5.25,0.35) {$\Scale[0.7]{J^+_0}$};
\node[] at (5.75,0.35) {$\Scale[0.7]{J^+_0}$};
\node[] at (6.25,0.35) {$\Scale[0.7]{J^+_0}$};

\draw[->, orange, line width=0.75] (4.78,0.06) -- (4.35,1.35);
\draw[->, orange, line width=0.75] (4.28,1.56) -- (3.85,2.85);

\node[] at (4.3,0.6) {$\Scale[0.7]{J^-_{-3}}$};
\node[] at (3.8,2.1) {$\Scale[0.7]{J^-_{-3}}$};

\end{tikzpicture}
\end{center}
\caption{The schematic form of the Verma module for a continuous representation with $w=2$ (left) and a discrete representation with $w=2$ (right). Here a black dot signifies the presence of at least one state in the Verma module with the indicated charges. When considering operators $V^w_{j,h}(x;y;z)$, the $x$ coordinate resums states on the magenta line, while the $y$ coordinates resums states on the cyan line. Notice that the right picture can be equivalently interpreted as the Verma module of a $\mathcal D^+$ representation with $w=2$ or as the Verma module of a $\mathcal D^-$ representation with $w=3$, which is a manifestation of eq.~\eqref{eq:identification representations D+ D-}.} \label{fig:Verma module}
\end{figure}
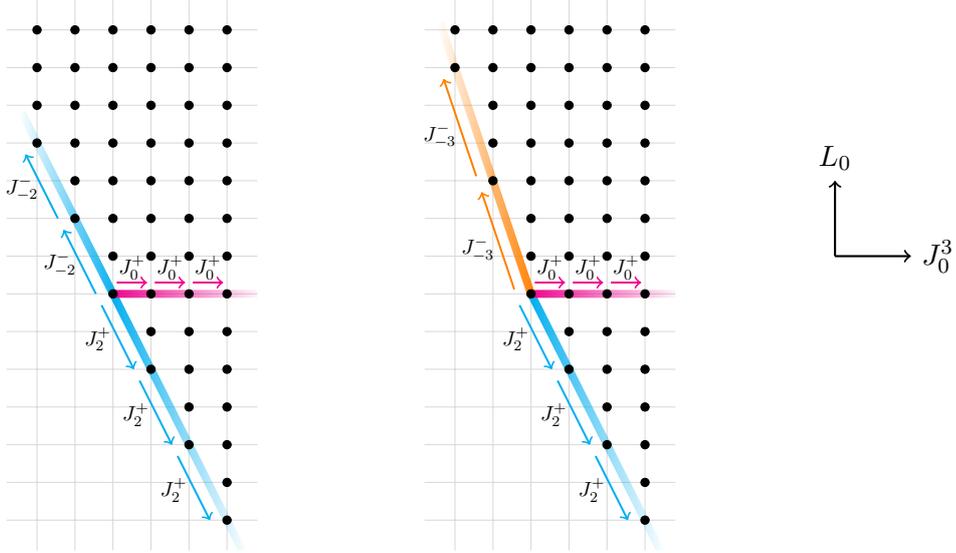

The natural arena to study holographic string correlators is the $x$-basis. From the worldsheet point of view, $x$ is an auxiliary variable that captures the transformation behaviour of the zero modes $J_0^a$. We will go one step further and introduce a \emph{third} position variable for each vertex operator that we call $y$. $y$ plays the same role for $J_w^+$, $J_0^3$ and $J_{-w}^-$ (i.e.~the image of the zero modes under spectral flow) as $x$ does for the zero modes $J_0^a$ (and the worldsheet coordinate $z$ does for the M\"obius generators $L_1$, $L_0$ and $L_{-1}$). To orient the reader, we have drawn in  Figure~\ref{fig:Verma module} the Verma module of a spectrally flowed continuous representation in the state picture. Every dot in the picture represents a state up to degeneracy. When going to the vertex operators, we resum the action of the three modes $J_0^+$, $J_w^+$ and $L_{-1}$. So formally the `operator state correspondence' gets extended by a third variable (we suppress right-movers):\footnote{An operator state correspondence of this form only exists for discrete representations $\mathcal{D}^+_j$. This is also the case for the usual operator state correspondence in CFTs, where one defines representations with bounded spectrum from below by translation of the state at 0. Nonetheless we can define these vertex operators for arbitrary representations. See eqs.~\eqref{y-basis}, \eqref{y-basis-discrete-plus} and \eqref{y-basis-discrete}.}
\be 
V_j^w(x;y;z)\equiv \mathrm{e}^{z L_{-1}}\mathrm{e}^{x J_0^+} \mathrm{e}^{y J_w^+} [\ket{j,j}]^{(w)}]\ . \label{eq:y basis operator state correspondence}
\ee
Note that $L_{-1}$ and $J_w^+$ don't commute, so the order in this definition is important. In Figure~\ref{fig:Verma module} we schematically indicated in cyan and magenta respectively the $J^+_w$  and $J^+_0$ resummed directions. As a consequence these vertex operators don't have a definite conformal weight on the worldsheet and in target space.
Similarly, one can write down an analogue of the translation identity \eqref{x-basis-vertex-op} of vertex operators to account also for the presence of $y$-space. 

The reason for introducing this auxiliary variable is mostly technical convenience. It turns out that we can give simple formulas for correlation functions using this $y$-basis. The corresponding expressions for the physical $h$-space are much more complicated. We are not aware of any physical interpretation of the $y$-variable in the context of holography.

Before discussing the meaning of this definition further, let us make some basic observations that will play an important role. For illustration, let us discuss the $x$-dependence of the unflowed three-point function that takes the familiar form
\be 
(x_1-x_2)^{-j_1-j_2+j_3}(x_1-x_3)^{-j_1-j_3+j_2}(x_2-x_3)^{-j_2-j_3+j_1}\ ,
\ee
where $x_1$, $x_2$ and $x_3$ are the three insertion points of the vertex operators in spacetime. While this is of course completely standard in CFT, it is at first sight puzzling from the state picture. Would we not expect a different three-point function depending on whether we consider a three-point function of the form $\langle \mathcal{D}^+_{j_1}\mathcal{D}^+_{j_2}\mathcal{D}^+_{j_3}\rangle $, $\langle \mathcal{D}^+_{j_1}\mathcal{C}_{j_2}\mathcal{D}^-_{j_3}\rangle$ etc? The answer is that all these three-point functions are united in the $x$-basis representation into this single formula and from this formula one can extract the answer in the state picture for every choice of representation. Essentially, the discrete representation $\mathcal{D}_j^+$ for the $i$-th field is obtained by expanding the three-point function around $x_i=0$. This is what one has in mind when writing the definition \eqref{eq:y basis operator state correspondence}. However, the $\mathcal{D}_j^-$ representation can be extracted similarly by expanding around $x_i=\infty$. Both these prescriptions lead manifestly to truncating power series in $x_i$ and they can be understood as contour integrals in the $x_i$-plane, either around 0 or $\infty$. To extract the continuous representation, one instead integrates over the full $x_i$-space (non-holomorphically). We hope that this will become clearer in the following.
For now it is sufficient to keep in mind that we will produce only a single formula for the $y$-basis correlators and from there extract continuous and discrete representations. 

The OPEs \eqref{eq:x-OPEs} take the following form in the $y$-basis:
\begin{subequations} \label{eq:xy-OPEs}
\begin{align}
J^+(\zeta) V^w_j(x;y;z)&=\sum_{p=1}^{w+1} \frac{(J^+_{p-1} V^w_j)(x;y;z)}{(\zeta-z)^p}+\mathcal{O}(1)\ , \\
\big(J^3(\zeta)-x J^+(\zeta) \big) V^w_j(x;y;z)&=\frac{(j+\frac{k w}{2}+y\partial_y) V^w_j(x;y;z)}{\zeta-z}+\mathcal{O}(1)\ , \\
\big(J^-(\zeta)-2 x J^3(\zeta) + x^2 J^+(\zeta)\big) V^w_j(x;y;z)&=(\zeta-z)^{w-1} (2jy+y^2 \partial_y) V^w_j(x;y;z) \nonumber \\
& \qquad +\mathcal{O}((\zeta-z)^{w})\ .
\end{align}
\end{subequations}
The shifts on the left-hand-side of these OPEs can be derived from \eqref{x-basis-vertex-op}. From the previous discussion we also have the identifications
\begin{subequations}
\begin{align}
(J_0^+ V^w_j)(x;y;z)&=\partial_x V_j^w(x;y;z)\ , \label{eq:J0p action y basis}\\
(J_w^+ V^w_j)(x;y;z)&=\partial_y V^w_j(x;y;z)\ ,\label{eq:Jwp action y basis}
\end{align}
\end{subequations}
Comparing these formulas with the unflowed sector OPEs \eqref{eq:x-OPEs unflowed}, we see that in the unflowed sector, the variables $x$ and $y$ (and the corresponding spins $h$ and $j$) coincide. This follows also directly from the definition \eqref{eq:y basis operator state correspondence}. We make the standard choice and keep $x$ and $j$ in the unflowed sector.

\paragraph{Continuous representations.} For continuous spectrally flowed  representations, the $\mathrm{e}^{y J_w^+}$ in \eqref{eq:y basis operator state correspondence} should be understood as follows,\footnote{The normalisation factor $\frac{i}{(2\pi)^2}$ in the right-hand-side of \eqref{y-basis} is just a convention.}
\be
V_{j}^w(x;y;z) \equiv \frac{i}{(2\pi)^2}\sum_{h - \bar h} \int_{-\infty}^\infty \mathrm{d}(h+\bar h) \, V_{j,h,\bar h}^w(x;z) \, y^{h-j-\frac{kw}{2}} \,  \bar y^{\bar h-j-\frac{kw}{2}} \ . 
\label{y-basis}
\ee
Note that we understand a continuous representation here as a sum of irreducible representations that would each be labelled by $h \mod \mathds{Z}$. This is convenient because in the $y$-basis they are all mixed together. 
In order for correlators to be single-valued, in eq.~\eqref{y-basis} the difference of holomorphic and antiholomorphic dual conformal dimensions takes value in $h-\bar h \in \mathds{Z}$. As follows from the integral identity \eqref{eq:two-singular-points}, the inverse transform is given by
\be 
V_{j,h,\bar{h}}^w(x;z)=  \int \mathrm{d}^2 y \, y^{\frac{kw}{2}+j-h-1} \, \bar{y}^{\frac{kw}{2}+j-\bar{h}-1} \, V_j^w(x;y;z) \ .
\label{eq:inverse-y-transform}
\ee
It is useful to also give a formal prescription of how this definition reduces to the unflowed sector,
\be 
V^0_{j,h,\bar h}(x;z)=\delta_{h,\bar h} \, \delta(h+\bar h-2j) \, V^0_j(x;y;z) \ , \qquad V^0_j(x;y;z) = V^0_j(x;z) \ . 
\label{y-transform-zero-spectral-flow}
\ee
The choice for the reflection coefficient made in \eqref{hmodZ-reflection-coeff} behaves nicely with respect to the $y$-transform. In fact, 
\be 
V^w_{1-j}(x;y;z) = R_{1-j} \frac{(1-2j)}{\pi} \int d^2 y' \, |y-y'|^{4j-4} \,  V^w_{j}(x;y';z) \ ,
\label{eq:y-basis-reflection}
\ee
which is the $y$-transform analogue of \eqref{eq:Teschner-reflection} for flowed representations. Eq.~\eqref{eq:y-basis-reflection} can be verified by substituting \eqref{y-basis} into the right-hand-side and then making use of the integral identity \eqref{eq:three-singular-points}. 

\paragraph{Discrete representations.} For discrete representations, the integral in \eqref{y-basis} is not necessary. The definition \eqref{eq:y basis operator state correspondence} reads instead
\begin{align}
& \mathcal{D}_j^+ :  & & V_{j}^w(x;y;z) \equiv \sum_{h, \bar h \, \in \, \mathds{Z}_{\geq 0} +j+\frac{kw}{2}}  \, V_{j,h,\bar h}^w(x;z) \, y^{h-j-\frac{kw}{2}} \,  \bar y^{\bar h-j-\frac{kw}{2}} \ , \label{y-basis-discrete-plus} \\
& \mathcal{D}_j^- :  & & V_{j}^w(x;y;z) \equiv \sum_{h, \bar h \, \in \, \mathds{Z}_{\leq 0} -j+\frac{kw}{2}}  \, V_{j,h,\bar h}^w(x;z) \, y^{h-j-\frac{kw}{2}} \,  \bar y^{\bar h-j-\frac{kw}{2}} \ , 
\label{y-basis-discrete}
\end{align}
The inverse transform reads
\begin{align}
& \mathcal{D}_j^+ : & &  V_{j,h,\bar{h}}^w(x;z)= \frac{1}{(2\pi i)^2} \oint_0 \mathrm{d} y \, \oint_0 \mathrm{d} \bar y \, y^{\frac{kw}{2}+j-h-1} \, \bar{y}^{\frac{kw}{2}+j-\bar{h}-1} \, V_j^w(x;y;z) \ ,
\label{eq:y-inverse-D+} \\
& \mathcal{D}_j^- : & &  V_{j,h,\bar{h}}^w(x;z)= \frac{1}{(2\pi i)^2} \oint_\infty \mathrm{d} y \, \oint_\infty \mathrm{d} \bar y \, y^{\frac{kw}{2}+j-h-1} \, \bar{y}^{\frac{kw}{2}+j-\bar{h}-1} \, V_j^w(x;y;z) \ , \label{eq:y-inverse-D-}
\end{align}
It follows directly from the definition \eqref{eq:y basis operator state correspondence} that the relevant contour in the case of $\mathcal{D}^+_j$ should run around 0, since this is the point where we identify the state and the vertex operator. We will also see that correlators of $y$-transformed fields behave as $\mathcal{O}(y^{-2j})$ for $y \to \infty$ and as $\mathcal{O}(1)$ for $y \to 0$. The integration around $0$ (respectively $\infty$) then implements the truncation $h-\frac{kw}{2} \geq j$ (respectively $h-\frac{kw}{2} \leq -j$), typical of lowest weight (respectively highest weight) discrete representations. 

\section{Global and local Ward identities}
\label{sec:constraints}

In this section we study the various constraints that correlators of the form
\be 
\left\langle \prod_{l=1}^n V_{j_l,h_l}^{w_l}(x_l;z_l)  \right \rangle 
\label{eq:generic-correlator}
\ee 
should obey. We mostly focus on symmetry constraints that apply equally to left- and right-movers. Consequently we shall often suppress the anti-holomorphic dependence of these correlators.
We start discussing global Ward identities on the worldsheet and in spacetime in Section \ref{sec:global-ward-id} and proceed in Section \ref{sec:local-ward-id} by reviewing local Ward identity constraints, mainly following \cite{Eberhardt:2019ywk}.  

\subsection{Global Ward identities}
\label{sec:global-ward-id}

Correlators satisfy global Ward identities both on the worldsheet and in spacetime. They respectively amount to 
\begin{subequations}
\begin{align}
& \sum_{i=1}^n \left(x_i^{\ell+1} \partial_{x_i}+(\ell +1) \, h_i \, x_i^\ell \right)\left\langle \prod_{l=1}^n V_{j_l,h_l}^{w_l}(x_l;z_l) \right \rangle=0\ , \label{eq:x-Ward-id} \\
& \sum_{i=1}^n \left(z_i^{\ell+1} \partial_{z_i} + (\ell+1) \ \Delta_i \ z_i^\ell \right) \left\langle \prod_{l=1}^n V_{j_l,h_l}^{w_l}(x_l;z_l) \right\rangle=0\ , \label{eq:z-Ward-id}
\end{align}
\label{eq:Ward-id}
\end{subequations}
for $\ell \in \{-1, 0, 1 \}$. Here
\be 
\Delta_i = -\frac{j_i \, (j_i-1)}{k-2}-w_i\,  h_i+\frac{k \, w_i^2}{4} 
\label{eq:Delta}
\ee
is the worldsheet conformal weight. See e.g.\ \cite{Maldacena:2000hw} for an explanation of this formula.
Similar relations hold for the anti-holomorphic coordinates $\bar x_i$ and $\bar z_i$. In terms of the $y$-transform introduced in Section \ref{sec:y-transform}, eqs.~\eqref{eq:Ward-id} read
\begin{subequations} 
\begin{align} 
& \sum_{i=1}^n \left(x_i^{\ell+1} \partial_{x_i}+(\ell+1) \, x_i^\ell \, \left(h_i^0+y_i \partial_{y_i}\right)\right) 
\left\langle \prod_{l=1}^n V_{j_l}^{w_l}(x_l;y_l;z_l) \right \rangle=0\ ,
\label{eq:x-Ward-id-ybasis} \\
& \sum_{i=1}^n \left(z_i^{\ell+1} \partial_{z_i}+(\ell+1) \, z_i^\ell \, \left(\Delta_i^0-w_i \, y_i \partial_{y_i}\right)\right) 
\left\langle \prod_{l=1}^n V_{j_l}^{w_l}(x_l;y_l;z_l) \right \rangle=0\ ,
\label{eq:z-Ward-id-ybasis}
\end{align}
\label{eq:xz-Ward-id-ybasis}
\end{subequations}
where again $\ell \in \{-1, 0, 1 \}$ and we defined
\begin{subequations}
\begin{align} 
h_i^0&=j_i+\frac{k \, w_i}{2} \ ,
\label{h^0} \\
\Delta_i^0&=-\frac{j_i \, (j_i-1)}{k-2}-w_i \left(j_i+\frac{k \, w_i}{4}\right)\ .
\label{Delta^0}
\end{align}
\end{subequations}
For two- and three-point functions global Ward identities specify the $x$- and $z$-dependence of the correlation function uniquely. We will thus often put the vertex operators at $x_1=z_1=0$, $x_2=z_2=\infty$ in the case of two-point functions and at $x_1=z_1=0$, $x_2=z_2=1$ and $x_3=z_3=\infty$ for three-point functions. The same is of course true for the anti-holomorphic part. 

Setting $x=\infty$ for the $y$-transform  translates to 
\be
V_j^w(\infty;y;z)\equiv \lim_{x\to \infty} x^{2 h^0} \, V_j^w(x;y \, x^2;z) \ . 
\ee
This follows from
\begin{align}
V_{j,h}^w(\infty;z) = \lim_{x \to \infty} x^{2 h} \, V_{j,h}^w(x;z) = \lim_{x \to \infty} x^{2h^0} \int \mathrm{d}^2y \ y^{\frac{kw}{2}+j-h-1} \ \bar y^{\frac{kw}{2}+j-\bar h-1} V_{j}^w(x;y \, x^2;z) \ ,  
\end{align}
where we performed the change of variables $y \to yx^2$ in the last equality and $h^0$ is given by eq.~\eqref{h^0}. Similarly, we can take $z$ to $\infty$ and obtain
\be 
V_j^w(x;y;\infty) \equiv \lim_{z \to \infty} z^{2 \Delta^0}  \, V_j^w(x;y \, z^{-2w};z) \ ,
\ee
where $\Delta^0$ is given in eq.~\eqref{Delta^0}. For two-point functions, the solution to the global Ward identities \eqref{eq:x-Ward-id-ybasis} and \eqref{eq:z-Ward-id-ybasis} reads
\be 
\langle V_{j_1}(x_1;y_1;z_1)V_{j_2}(x_2;y_2;z_2) \rangle=x_{12}^{-h_1^0-h_2^0}z_{12}^{-\Delta_1^0-\Delta_2^0}F_2\left(\frac{y_1z_{12}^{w_1}}{x_{12}},\frac{y_2z_{12}^{w_2}}{x_{12}}\right)\ . \label{eq:global Ward identities solution 2pt function}
\ee
where $x_{ij}=x_i-x_j$. 
The function $F_2$ satisfies furthermore the following two differential equations
\begin{subequations} \label{eq:2pt function global Ward identities constraints}
\begin{align}
0&=(h_1^0-h_2^0)F_2(y_1,y_2)+(y_1 \partial_{y_1}-y_2 \partial_{y_2} ) F_2(y_1,y_2)\ , \\
0&=(\Delta_1^0-\Delta_2^0)F_2(y_1,y_2)-(w_1 y_1 \partial_{y_1}-w_2 y_2 \partial_{y_2} ) F_2(y_1,y_2)\ . 
\end{align}
\end{subequations}
For three-point functions we have instead
\begin{multline} 
\langle V_{j_1}(x_1;y_1;z_1)V_{j_2}(x_2;y_2;z_2)V_{j_3}(x_3;y_3;z_3) \rangle  \\
= \prod_{i<j} x_{ij}^{h_\ell^0-h_i^0-h_j^0}z_{ij}^{\Delta_\ell^0-\Delta_i^0-\Delta_j^0} F_3\left(\left\{y_i \, \frac{x_{i+1,i+2} \, z_{i,i+1}^{w_i} \, z_{i,i+2}^{w_i}}{x_{i,i+1} \, x_{i,i+2} \, z_{i+1,i+2}^{w_i}}\right\}_{i=1,2,3}\right) \ ,  \label{eq:global Ward identities solution 3pt function}
\end{multline}
where $\ell$ is the third index not equal to $i$ and $j$, the indices are understood to be mod 3.

\subsection{Local Ward identities}
\label{sec:local-ward-id}

As discussed at length in \cite{Eberhardt:2019ywk}, in addition to \emph{global} Ward identities, also \emph{local} Ward identities can be derived for the correlator \eqref{eq:generic-correlator}. As it turns out, local Ward identities imply a set of recursion relations in the spacetime conformal dimension $h_l$. For two-point functions with $w_1=w_2=w>0$, they take the form 
\begin{multline}
\frac{(z_1-z_2)^w}{x_1-x_2} \left(h_1-\tfrac{k \, w}{2}+j_1-1\right)\left\langle h_1-1 \right \rangle_2 = \frac{x_2-x_1}{(z_2-z_1)^w} \left(h_2-\tfrac{k \, w}{2}+1-j_2\right)\left\langle h_2+1 \right \rangle_2 \\
+(h_1-h_2)\left\langle \, \dots \, \right \rangle_2 \ ,
\label{eq:two-pt-function-recursion}
\end{multline}
where we have introduced the short-hand notation 
\begin{subequations} 
\begin{align}
\left\langle \, \dots \, \right \rangle_n & \equiv \left\langle \prod_{l=1}^n V_{h_l,\bar{h}_l,j_l}^{w_l}(x_l;z_l) \right \rangle \ , \\
\left\langle h_p + r \right \rangle_n & \equiv \left\langle V_{h_p+r,\bar{h}_p,j_p}^{w_p}(x_p; z_p) \prod_{l\neq p} V_{h_l,\bar{h}_l,j_l}^{w_l}(x_l;z_l) \right \rangle \ . 
\end{align}
\label{eq:two-point-dots}
\end{subequations}
An additional local Ward identity can be obtained by exchanging the roles of $x_1, x_2$, $h_1, h_2$ and $j_1,  j_2$ in \eqref{eq:two-pt-function-recursion}. 

In a similar fashion, recursion relations in shifted values of the $h_l$ can be derived for $3$-point functions as well, provided that\footnote{Correlators that violate \eqref{eq:w-bound1}, vanish identically \cite{Maldacena:2001km,Eberhardt:2019ywk}. }
\be 
\sum_{i \neq j} w_i \geq w_j -1 \qquad \text{for all } j \in \{1,2,3 \} \ . 
\label{eq:w-bound1}
\ee
For example, when $\sum_i w_i \in 2 \mathds{Z}+1$, $w_i > 0$ and $w_i+w_j \geq w_k +1$ for all $(i,j,k)$, three-point function recursion relations read \cite{Eberhardt:2019ywk}\footnote{We choose the insertion points as $x_1=z_1=0$, $x_2=z_2=1$ and $x_3=z_3= \infty$.}
\begin{multline}
a_i^{-1} \left(h_i-\tfrac{k}{2}w_i+j_i-1 \right) \left\langle h_i-1 \right \rangle_3 = \left(\frac{w_i}{N}- 1\right)a_i \left(h_i-\tfrac{k}{2}w_i-j_i+1 \right) \left\langle h_i+1 \right \rangle_3\\
\hspace{100pt} + \frac{w_{i+1}}{N} a_{i+1} \left(h_{i+1}-\tfrac{k}{2}w_{i+1}-j_{i+1}+1 \right)\left\langle h_{i+1}+1 \right \rangle_3\\
\hspace{100pt} + \frac{w_{i+2}}{N} a_{i+2} \left(h_{i+2}-\tfrac{k}{2}w_{i+2}-j_{i+2}+1 \right) \left\langle h_{i+2}+1 \right \rangle_3 \\
+\left(-\frac{w_i}{N}\sum_{l=1}^n h_l +2h_i\right)  \left\langle \, \dots \, \right \rangle_3 \ , 
\label{eq:three-pt-function-recursions}
\end{multline}
where indices are understood to be mod 3, $N=\frac{1}{2}\sum_i (w_i-1)+1$ and
\be 
a_i=\frac{\begin{pmatrix} \frac{1}{2} (w_i+w_{i+1}+w_{i+2}-1) \\\frac{1}{2} (-w_i+w_{i+1}+w_{i+2}-1) \end{pmatrix}}{\begin{pmatrix} \frac{1}{2} (-w_i+w_{i+1}-w_{i+2}-1) \\ \frac{1}{2} (w_i+w_{i+1}-w_{i+2}-1)\end{pmatrix}}\ . \label{eq:three-pt-ai}
\ee
The $y$-transform of \eqref{eq:three-pt-function-recursions} will be useful in the following. It reads 
\begin{multline}
\Bigl[ -a_i^{-1} y_i \, (y_i \partial_{y_i} + 2j_i) + \left(\frac{w_i}{N}-1\right)a_i \, \partial_{y_i}+  \\
\frac{w_{i+1}}{N} \, a_{i+1} \,  \partial_{y_{i+1}} +\frac{w_{i+2}}{N} \, a_{i+2} \, \partial_{y_{i+2}} -\frac{w_i}{N} \sum_{l=1}^n \left(y_l \partial_{y_l}+j_l+\tfrac{k}{2}w_l \right) \\
+ 2 \left(y_i \partial_{y_i}+j_i+\tfrac{k}{2} w_i\right) \Bigr] \left\langle V_{j_1}^{w_1}(0; y_1;0) \, V_{j_2}^{w_2}(1; y_2;1) \, V_{j_3}^{w_3}(\infty; y_3; \infty) \right\rangle = 0 \ . 
\label{eq:y-recursion-relations}
\end{multline}
The fact that these constraints become partial differential equations in $y$-space is the principal reason that we introduce the $y$-transform in the first place.
Similarly, for all the configurations of $w_i$ satisfying \eqref{eq:w-bound1}, recursion relations can be derived explicitly for each $w_i > 0$. There is no known analytic derivation of these constraints in closed form for all values of spectral flow. However, one can always derive them for a given choice of spectral flow $w$ by solving a linear system. The reader can find more details about their derivation in \cite{Eberhardt:2019ywk} and reproduce them explicitly with the ancillary {\tt Mathematica} file.  

\paragraph{Truncations.} Let us comment on the representation coefficients in \eqref{flowed-rep}. While this choice is legitimate --- it gives rise to the Casimir \eqref{casimir} --- some readers may find it unnatural and may prefer to flip the spins according to $j \to 1-j$. The choice made in \eqref{flowed-rep} is however necessary in order to implement the correct truncation of discrete representations at the level of local Ward identities. Let us explain this by means of an example. Consider the local Ward identities of \cite{Eberhardt:2019ywk} for a three-point function with spectral flow parameters $(w_1, w_2, w_3)=(1,1,1)$,  
\begin{multline}
c_{j_1}^- \left\langle V^1_{j_1,h_1-1}(0;0) V^1_{j_2,h_2}(1;1) V^1_{j_3,h_3}(\infty;\infty) \right\rangle\\
-(h_1-h_2-h_3)\left\langle V^1_{j_1,h_1}(0;0) V^1_{j_2,h_2}(1;1) V^1_{j_3,h_3}(\infty;\infty) \right\rangle \\
- c_{j_3}^+ \left\langle V^1_{j_1,h_1}(0;0) V^1_{j_2,h_2}(1;1) V^1_{j_3,h_3+1}(\infty;\infty) \right\rangle \\
= c_{j_2}^+ \left\langle V^1_{j_1,h_1}(0;0) V^1_{j_2,h_2+1}(1;1) V^1_{j_3,h_3}(\infty;\infty) \right\rangle \ , 
\label{111-recursion}
\end{multline}
where $h_i = m_i+\frac{k}{2}$
and we defined the representation coefficients $c_j^\pm$ according to
\be 
J^+_{w} [\ket{j,m}]^w = c^{+}_j [\ket{j,m+1}]^w \ , \qquad J^-_{-w} [\ket{j,m}]^w = c^{-}_j [\ket{j,m-1}]^w \ . 
\ee
We shall consider the case where all vertex operators entering the correlator in the right-hand-side of \eqref{111-recursion} are lowest weight states of a $\mathcal{D}_j^+$ representation, see eq.~\eqref{DiscreteD+}. This corresponds to the choice
\be
h_1=\frac{k}{2}+j_1 \ ,  \qquad h_2 =-1+\frac{k}{2}+ j_2 \ , \qquad h_3 =\frac{k}{2} + j_3 \ . 
\ee
All correlators in the left-hand-side of \eqref{111-recursion} vanish, as they contain vertex operators with $h_2\le -1+\frac{k}{2}+ j_2$, which is outside the range \eqref{DiscreteD+}. It follows that $c_{j_2}^+$ must also vanish for $h_2 =-1+\frac{k}{2}+ j_2$. Hence a consistent choice is
\be 
c_{j}^+ = h -\frac{k}{2} +1- j \ , 
\ee
as in eq.~\eqref{flowed-rep}. The representation coefficient $c_j^-$ can be fixed by a similar argument or by requiring the Casimir to be as in eq.~\eqref{casimir}. On top of being the convention chosen by some of the standard references in the literature \cite{Maldacena:2000hw, Maldacena:2000kv, Maldacena:2001km}, it will become clear in the next sections that this choice of representation coefficients also makes all formulas more uniform.

\section{Two-point functions}
\label{sec:2ptf}

As a warm up to three-point functions in the next section, we consider two-point functions first,
\be 
\left\langle V^{w_1}_{j_1, h_1}(x_1; z_1) \,  V^{w_2}_{j_2, h_2}(x_2; z_2) \right\rangle
\label{eq:2ptf}
\ee
with $w_1, w_2 >0$. We should mention that the following derivation is not necessarily the quickest route to the two-point function \eqref{eq:2ptf}: the aim here is to introduce the reader to Section \ref{sec:3ptf}, where three-point functions will be studied by similar techniques. We can be completely explicit in all the computations and show that \eqref{eq:2ptf} is entirely fixed by symmetries up to normalization. These conditions will be stronger than the solution of the global Ward identities \eqref{eq:global Ward identities solution 2pt function}. For instance they imply that the two-point function can be only non-vanishing for $w_1=w_2$. Let us look at the correlator with the insertion of a $J^a(z)$ current, 
\be 
\left\langle J^a(z) \, V^{w_1}_{j_1, h_1}(x_1; z_1) \,  V^{w_2}_{j_2, h_2}(x_2; z_2) \right\rangle \ . 
\ee
One can derive a global Ward identity for this type of correlator as well, which relates the three correlators with $J^+(z)$, $J^3(z)$ and $J^-(z)$ insertions. Solvability of these constraints requires
\be 
h_1 = h_2 \qquad \text{or} \qquad h_1 = h_2+1 \qquad \text{or} \qquad h_1 = h_2 -1 
\label{eq:two-ptf-h1-h2}
\ee 
and fixes the coordinate dependence together with the worldsheet global Ward identities to be 
\begin{subequations}
\begin{align}
& \left\langle J^+(z) \, V^{w_1}_{j_1, h_1}(x_1; z_1) \,  V^{w_2}_{j_2, h_2}(x_2; z_2) \right\rangle \nonumber \\
& \quad  = \frac{C \, (x_1-x_2)^{-h_1-h_2-1}}{(z-z_1)^{\Delta_1-\Delta_2+1}(z-z_2)^{\Delta_2-\Delta_1+1}(z_1-z_2)^{\Delta_1+\Delta_2-1}} \ , \\
& \left\langle J^3(z) \, V^{w_1}_{j_1, h_1}(x_1; z_1) \,  V^{w_2}_{j_2, h_2}(x_2; z_2) \right\rangle \nonumber \\
& \quad = \frac{(1-h_1+h_2)\, x_1 + (1+h_1-h_2) \, x_2}{2} \, \left\langle J^+(z) \, V^{w_1}_{j_1, h_1}(x_1; z_1) \,  V^{w_2}_{j_2, h_2}(x_2; z_2) \right\rangle \ ,  \\
& \left\langle J^-(z) \, V^{w_1}_{j_1, h_1}(x_1; z_1) \,  V^{w_2}_{j_2, h_2}(x_2; z_2) \right\rangle \nonumber \\
& \quad = \frac{(h_1-h_2)(h_1-h_2-1)\, x_1^2 - 2((h_2-h_1)^2-1)\, x_1x_2 + (h_1-h_2)(h_1-h_2+1)\, x_2^2}{2} \nonumber  \\
& \hspace{50pt} \times \, \left\langle J^+(z) \, V^{w_1}_{j_1, h_1}(x_1; z_1) \,  V^{w_2}_{j_2, h_2}(x_2; z_2) \right\rangle \ ,
\end{align}
\label{eq:two-ptf-with-J} 
\end{subequations}
where $C$ is an arbitrary constant that may depend on $w_1, w_2, j_1, j_2, h_1, h_2$. In the following we assume eq.~\eqref{eq:two-ptf-h1-h2} to be always satisfied. The analysis of \cite{Eberhardt:2019ywk} suggests to consider for $z \to z_i$ the combination
\begin{multline}
\left\langle \bigl(J^-(z)-2 \,  x_i \,  J^3(z)+x_i^2 J^+(z)\bigr) \, V^{w_1}_{j_1, h_1}(x_1; z_1) \,  V^{w_2}_{j_2, h_2}(x_2; z_2) \right\rangle \\
= (z-z_i)^{w_i-1} \Bigl(h_i-\frac{k \, w_i}{2}+j_i-1 \Bigr) \left\langle h_i-1 \right\rangle_2  + \mathcal{O}\bigl((z-z_i)^w \bigr)\ , 
\label{eq:two-point-recursion-derivation}
\end{multline}
where the right-hand-side follows from the form of the OPE \eqref{regular-combination-OPE} and \eqref{eq:Jwp action} and we used the notation \eqref{eq:two-point-dots}. On the other hand, making use of \eqref{eq:two-ptf-h1-h2} and \eqref{eq:two-ptf-with-J}, one calculates for the left-hand-side of this equation
\be 
=\frac{1}{2} \left(x_1-x_2\right)^2 \left(h_i-h_j\right) \left(h_i-h_j+1\right)\left\langle J^+(z) \, V^{w_1}_{j_1, h_1}(x_1; z_1) \,  V^{w_2}_{j_2, h_2}(x_2; z_2) \right\rangle\ ,
\ee
where $j$ is the index not equal to $i$. Hence the left-hand-side of eq.~\eqref{eq:two-point-recursion-derivation} vanishes identically whenever $h_i-h_j \in \{ -1,0 \}$.
Comparing with eq.~\eqref{eq:two-point-recursion-derivation}, this implies that also $\langle h_i-1 \rangle=0$ whenever $h_i-h_j \in \{-1,0\}$. This means it can only be non-vanishing for the second possibility in eq.~\eqref{eq:two-ptf-h1-h2} if $i=1$ and for the last possibility if instead $i=2$. 
Let us now consider the case in which the left-hand-side of \eqref{eq:two-point-recursion-derivation} does not vanish identically. This happens for example when in \eqref{eq:two-point-recursion-derivation} we have $i=1$ and $h_2 = h_1-1$. In this case, it follows from \eqref{eq:two-ptf-with-J} that 
\begin{multline}
\left\langle \bigl(J^-(z)-2 \,  x_i \,  J^3(z)+x_i^2 J^+(z)\bigr) \, V^{w_1}_{j_1, h_1}(x_1; z_1) \,  V^{w_2}_{j_2, h_2}(x_2; z_2) \right\rangle \\
= \frac{C \, (x_1-x_2)^{-2h_1+2}}{(z-z_1)^{\Delta_1-\Delta_2+1}(z-z_2)^{\Delta_2-\Delta_1+1}(z_1-z_2)^{\Delta_1+\Delta_2-1}} \ . 
\label{eq:2ptf-z-z1-power}
\end{multline}
By comparing the power of $(z-z_1)$ in \eqref{eq:2ptf-z-z1-power} and \eqref{eq:two-point-recursion-derivation} we find 
\be 
w_1 - 1 = -1 -\Delta_1 + \Delta_2 \ .  
\label{eq:2ptf-match-z-z1-power}
\ee
Notice that $\Delta_i$ depends on $h_i$ through \eqref{eq:Delta} and eq.~\eqref{eq:2ptf-match-z-z1-power} must hold as a polynomial equation in $h_1=h_2+1$. It follows that $w_1 =w_2$,  which in turn implies $j_1 = j_2$ or $j_1 = 1-j_2$. To summarize, the two-point function \eqref{eq:2ptf} takes the form
\begin{multline}
\left\langle V^{w_1}_{j_1, h_1}(x_1; z_1) \,  V^{w_2}_{j_2, h_2}(x_2; z_2) \right\rangle = \delta(h_1-h_2) \, \delta_{w_1,w_2} \ (x_1-x_2)^{-h_1} (x_2-x_1)^{-h_1} \, \times \\ 
\times \,  (z_1-z_2)^{-\Delta_1} (z_2 -z_1)^{-\Delta_1} \Bigl( A \, \delta(j_1-j_2) + B \, \delta(j_1+j_2-1) \Bigr) \ ,
\label{eq:2ptf-general-form}
\end{multline}
where $A$ and $B$ are arbitrary constants that may depend on $h_1, j_1, w_1$ and $k$. The $h_i$ dependence of $A$ and $B$ can be fixed by considering the recursion relation \eqref{eq:two-pt-function-recursion}, that together with \eqref{eq:2ptf-general-form} imply
\be 
A = \text{const. } \binom{h_1-\frac{kw_1}{2}+j_1-1}{2j_1-1} \ , \qquad B = \text{const. } \ , 
\label{eq:AB-2ptf}
\ee
where the remaining constants depend just on $j_1, j_2,k$ and possibly on the fractional part of $h_1$. The constants $A$ and $B$ in \eqref{eq:2ptf-general-form} are purely conventional and for continuous representations their ratio is related to the choice of reflection symmetry coefficient.

\paragraph{Normalization.} We will now fix the normalization of the vertex operators. We choose $B \equiv (2 \pi)^2 i$. Reflection symmetry then also fully determines $A$, so that for continuous representations we obtain
\begin{multline}
\left\langle V^{w_1}_{j_1, h_1}(x_1; z_1) \,  V^{w_2}_{j_2, h_2}(x_2; z_2) \right\rangle =(2 \pi)^2 i \,   \delta^{(2)}(h_1-h_2) \, \delta_{w_1,w_2} \ (x_1-x_2)^{-h_1} (x_2-x_1)^{-h_1} \\ 
\times (\bar{x}_1-\bar{x}_2)^{-\bar{h}_1} (\bar{x}_2-\bar{x}_1)^{-\bar h_1}  (z_1-z_2)^{-\Delta_1} (z_2 -z_1)^{-\Delta_1}(\bar{z}_1-\bar{z}_2)^{-\bar{\Delta}_1} (\bar{z}_2 -\bar{z}_1)^{-\bar \Delta_1}\\
\times \Bigl( R(j_1,h_1,\bar h_1) \, \delta(j_1-j_2) + \delta(j_1+j_2-1) \Bigr) \ . 
\label{eq:2ptf-final}
\end{multline}
Here, $\delta^{(2)}(h)\equiv \delta_{h,\bar{h}} \delta(h+\bar{h})$. The symmetrization in the indices 1 and 2 is necessary to get the correct signs of the result. The imaginary unit $i$ in front might seem strange. It comes from requiring a real two-point function in the $y$-basis where this result takes the form (putting $x_1=z_1=0$ and $x_2=z_2=\infty$ for simplicity):
\begin{multline}
\left\langle V^{w_1}_{j_1}(0;y_1;0) \,  V^{w_2}_{j_2}(\infty;y_2;\infty) \right\rangle =\delta_{w_1,w_2}\,  B(j_1) \delta(j_1-j_2) \left|1- y_1 y_2 \right|^{-4j_1}\\
+\delta_{w_1,w_2} \, \delta(j_1+j_2-1) |y_1|^{-2j_1}|y_2|^{-2j_2} \delta^{(2)} \left(1- y_1 y_2 \right)\ .
\label{eq:2ptf y basis}
\end{multline}
Eq.~\eqref{eq:2ptf y basis} follows from local and global Ward identities in the $y_1$, $y_2$ variables. Here, $B(j)$ also appears in the unflowed two-point function \eqref{eq:normalization vertex operators} and is related to the reflection coefficient. This formula is of course consistent with the global Ward identities \eqref{eq:global Ward identities solution 2pt function} and \eqref{eq:2pt function global Ward identities constraints}. This result is actually almost identical to the unflowed two-point function provided that we identify $y_1=x_1$ and $y_2=x_2^{-1}$. We will see something similar happening for three-point functions.
Eqs.~\eqref{eq:2ptf-general-form} and \eqref{eq:AB-2ptf} agree with eq.~(5.13) of \cite{Maldacena:2001km}, where the two-point function was computed by different techniques. Notice that eq.~\eqref{eq:2ptf-final} is manifestly reflection symmetric. In fact, without any need to use the recursion relations \eqref{eq:two-pt-function-recursion}, one could directly fix $A$ and $B$ in \eqref{eq:2ptf-general-form} by simply requiring consistency with refection symmetry. 

\section{Three-point functions}
\label{sec:3ptf}

Let us proceed with the analysis of the three-point functions 
\be 
\left\langle V^{w_1}_{j_1, h_1}(x_1; z_1) \,  V^{w_2}_{j_2, h_2}(x_2; z_2) \, V^{w_3}_{j_3, h_3}(x_3; z_3) \right\rangle \ . 
\label{eq:3ptf}
\ee
The local Ward identities of \cite{Eberhardt:2019ywk} imply that these correlators vanish unless\footnote{Strictly speaking, in \cite{Eberhardt:2019ywk} it was checked that correlators with all $w_i \geq 1$ vanish unless \eqref{eq:w-bound} is obeyed. We have checked extensively that this remains the case when the condition $w_i \geq 1$ is relaxed in order to allow for unflowed vertex operators. The same bound was obtained by different techniques in \cite{Maldacena:2001km}.}
\be 
\sum_{i \neq j} w_i \geq w_j -1 \qquad \text{for all } j \in \{1,2,3 \} \ . 
\label{eq:w-bound}
\ee
When the  bound \eqref{eq:w-bound} is obeyed, the $h$-dependence of the three-point functions \eqref{eq:3ptf} is constrained by the recursion relations discussed in Section \ref{sec:local-ward-id}. In this section we will show that these fix uniquely the $h$-dependence of the structure constants. We will make a very natural conjecture about the overall normalization of three-point functions.

\subsection[The \texorpdfstring{$y$}{y}-dependence of the three-point function]{The $\boldsymbol{y}$-dependence of the three-point function}
\label{sec:y-dependence-3ptf}

It is convenient to divide the discussion of the configurations that satisfy \eqref{eq:w-bound} accordingly to the parity of $\sum_{i} w_i$. We will refer to them as \emph{odd parity} and \emph{even parity} case respectively. For each case we consider the $y$-transform of eq.~\eqref{eq:3ptf}, 
\be 
\left\langle V^{w_1}_{j_1}(0; y_1; 0) \,  V^{w_2}_{j_2}(1; y_2; 1) \, V^{w_3}_{j_3}(\infty; y_3; \infty) \right\rangle \ , 
\label{eq:y-3pt}
\ee
where we used the global Ward identities \eqref{eq:xz-Ward-id-ybasis} to fix the insertion points on the worldsheet and in spacetime. Notice the technical advantage of introducing the $y$-transform: the recursion relations turn into a system of three linear first order partial differential equations. We have solved this system in { \tt Mathematica} for all choices of $(w_1,w_2,w_3)$ obeying \eqref{eq:w-bound} and $\sum_i w_i \leq 25$. These are 1092 possible choices for $(w_1,w_2,w_3)$. 
Our strategy is then to guess the general solution to the local Ward identities from the collected data. For the odd parity case we spot the following pattern (up to an overall constant and not including the right-movers), 
\begin{align}
\left\langle V^{w_1}_{j_1}(0; y_1; 0) \,  V^{w_2}_{j_2}(1; y_2; 1) \, V^{w_3}_{j_3}(\infty; y_3; \infty) \right\rangle = X_{123}^{\frac{k}{2}-j_1-j_2-j_3} \prod_{i=1}^3 X_i^{-\frac{k}{2}+j_1+j_2+j_3-2j_i} \ ,
\label{3pt-odd-parity}
\end{align}
while for the even parity case we find
\begin{align}
\left\langle V^{w_1}_{j_1}(0; y_1; 0) \,  V^{w_2}_{j_2}(1; y_2; 1) \, V^{w_3}_{j_3}(\infty; y_3; \infty) \right\rangle = X_\emptyset^{j_1+j_2+j_3-k}\prod_{i<\ell} X_{i \ell}^{j_1+j_2+j_3-2j_i-2j_\ell} \  ,
\label{3pt-even-parity}
\end{align}
where for any subset $I \subset \{ 1,2,3 \}$, we defined
\be 
X_I(y_1,y_2,y_3)= \sum_{i \in I:\ \varepsilon_i=\pm 1} P_{\boldsymbol{w}+\sum_{i \in I} \varepsilon_i e_{i}} \prod_{i\in I} y_i^{\frac{1-\varepsilon_i}{2}} \ . 
\label{X_I-3pt}
\ee
Notice that the prefactor $X_\emptyset^{j_1+j_2+j_3-k}$ is just a number and only leads to a normalization of the answer. We will explain its presence later.
In eq.~\eqref{X_I-3pt}, the spectral flow parameters are chosen as $\boldsymbol{w}=(w_1, w_2, w_3)$ and 
\be 
e_1 = (1,0,0) \ , \quad e_2 = (0,1,0) \ , \quad e_3 = (0,0,1) \ . 
\ee
The numbers $P_{\boldsymbol{w}}$ are defined as follows,  
\be 
P_{\boldsymbol{w}} = 0 \qquad \text{for} \qquad \sum_j w_j < 2 \max_{i=1,2,3} w_i \quad \text{or}\quad \sum_i w_i \in 2\mathds{Z}+1
\ee
and otherwise 
\be
P_{\boldsymbol{w}} =S_{\boldsymbol{w}} \frac{G\left(\frac{-w_1+w_2+w_3}{2} +1\right) G\left(\frac{w_1-w_2+w_3}{2} +1\right) G\left(\frac{w_1+w_2-w_3}{2} +1\right) G\left(\frac{w_1+w_2+w_3}{2}+1\right)}{G(w_1+1) G(w_2+1) G(w_3+1)}  \ , 
\label{Pw-definition}
\ee
where $G(n)$ is the Barnes G function, defined as
\be 
G(n)=\prod_{i=1}^{n-1} \Gamma(i)
\label{barnesG}
\ee
for positive integer values. See Appendix~\ref{app:special functions} for more details on this function.
The function $S_{\boldsymbol{w}}$ is a phase depending on $\boldsymbol{w} \bmod 2$. It is defined as
\be 
S_{\boldsymbol w} = (-1)^{\frac{1}{2} x(x+1)} \ ,  \label{eq:Pw signs}
\ee  
where 
\begin{subequations}
\begin{align}
\boldsymbol{w} & \sim (0,0,0): \quad x = \frac{w_1+w_2+w_3}{2} \ ,   & 
\boldsymbol{w} & \sim (1,0,1): \quad x = \frac{w_1+w_2-w_3}{2}  \ , 	\\
\boldsymbol{w} & \sim (0,1,1): \quad x = \frac{w_1-w_2+w_3}{2} \ ,  & 
\boldsymbol{w} & \sim (1,1,0): \quad x = \frac{-w_1+w_2+w_3}{2} \ .
\end{align}
\end{subequations}
For example (again up to an overall constant and without including right-movers)
\begin{align}
&\left\langle V_{j_1}^3(0;y_1;0)V_{j_2}^3(1;y_2;1)V_{j_3}^3(\infty;y_3;\infty) \right \rangle=\left(10-y_1\right){}^{-j_1+j_2+j_3-\frac{k}{2}}
   \left(10-y_2\right){}^{j_1-j_2+j_3-\frac{k}{2}} \nonumber\\
   &\quad\times\left(1-6
   y_3\right){}^{j_1+j_2-j_3-\frac{k}{2}} \left(20-y_1-y_2-20 y_3-4 y_1
   y_3-4 y_2 y_3+y_1 y_2 y_3\right){}^{\frac{k}{2}-j_1-j_2-j_3} \ , \\
 &\left\langle V_{j_1}^4(0;y_1;0)V_{j_2}^6(1;y_2;1)V_{j_3}^4(\infty;y_3;\infty) \right \rangle=  \left(35+y_1-50 y_2+10 y_1 y_2\right){}^{-j_1-j_2+j_3} \nonumber\\
 &\quad\times\left(175-15
   y_1+15 y_3+y_1y_3\right){}^{-j_1+j_2-j_3} \left(-35-50
   y_2+y_3-10 y_3 y_2\right){}^{j_1-j_2-j_3}\ ,
\end{align}
so the reader sees that these solutions are completely explicit. 

Let us pause for a moment to take stock and underline the simplicity of this result. We started out to solve a very complicated linear system with $w_1+w_2+w_3$ unknowns that enters the derivation of the local Ward identities. After eliminating the $w_1+w_2+w_3-3$ `trivial' unknowns \eqref{intro-unknowns} as in  \cite{Eberhardt:2019ywk}, we remain with three equations whose explicit form in a special case we have given in eq.~\eqref{eq:y-recursion-relations}. It is quite surprising that the solution of these equations turns out to be a product of simple factors, each linear in $y_i$, and raised to a power that only depends on the spins $j_i$ and $k$.\footnote{We should point out that our guess can actually be proven in the restricted range $\sum_i (w_i-1) \ge 2\max_i (w_i-1)$ for the odd parity case. In this case the methods of \cite{Eberhardt:2019ywk, Eberhardt:2020akk} give an analytic derivation of the local Ward identities \eqref{eq:y-recursion-relations} and one can check that the solution we have given indeed always solves these constraints.}
\paragraph{Exchange symmetry.} Let us also mention another cross check that we performed on our conjectured solution for the three-point functions. The worldsheet CFT that we consider is a bosonic CFT and as such three-point functions should have bosonic exchange symmetry. It is by no means obvious that our formulae \eqref{3pt-odd-parity} and \eqref{3pt-even-parity} satisfy this constraint. This requires that $P_{(w_1,w_2,w_3)}$ and $P_{(w_2,w_1,w_3)}$ agree up to signs and those signs determine the statistics of the interchange. One can check that the perhaps peculiar looking signs \eqref{eq:Pw signs} are exactly what is needed to give an answer that has bosonic exchange statistics.

\subsection[The \texorpdfstring{$h$}{h}-dependence of structure constants]{The $\boldsymbol h$-dependence of structure constants} \label{subsec:three point function h dependence}

Now that we know the explicit form of the three-point functions \eqref{eq:y-3pt}, solutions to the recursion relations can be obtained by simply applying the inverse $y$-transform \eqref{eq:inverse-y-transform} (resp.~\eqref{eq:y-inverse-D+} and \eqref{eq:y-inverse-D-} for discrete representations). For continuous representations, the solution to the recursion relations takes then the form
\be 
\int \prod_{i=1}^3 \mathrm{d}^2 y_i \, y_i^{\frac{kw_i}{2}+j_i-h_i-1} \, \bar{y_i}^{\frac{kw_i}{2}+j_i-\bar{h}_i-1} \left\langle V^{w_1}_{j_1}(0; y_1; 0) \,  V^{w_2}_{j_2}(1; y_2; 1) \, V^{w_3}_{j_3}(\infty; y_3; \infty) \right\rangle \ , 
\label{h-basis-3pt}
\ee
with the correlator in the right-hand-side given by eqs.~\eqref{3pt-odd-parity} and \eqref{3pt-even-parity} respectively for odd and even parity.

We have not managed to perform the integration in \eqref{h-basis-3pt} explicitly for every choice of the spectral flow parameters $w_i$. However, we have been able to find an explicit result for the restricted range
\be 
\sum_{i=1}^3 w_i \in 2\mathds{Z}_{\ge 1}+1\ \ \text{and}\ \ \sum_{i=1}^3 w_i >2 \max_{i} w_i \ .  \label{eq:3pt function covering map constraint}
\ee
In this case the $y$-transformed solution \eqref{3pt-odd-parity} can be rewritten as (up to a normalization that we will reinstate later)\footnote{This is not manifest at this stage, but it can be easily checked with the help of the ancillary {\tt Mathematica} notebook or by using the functional equation of the Barnes G-function.}
\begin{align}
& \hspace{-15pt} \left\langle V^{w_1}_{j_1}(0; y_1; 0) \,  V^{w_2}_{j_2}(1; y_2; 1) \, V^{w_3}_{j_3}(\infty; y_3; \infty) \right\rangle \nonumber \\
& = \prod_{i=1}^3 (1-t_i)^{j_1+j_2+j_3-2j_i-\frac{k}{2}} \nonumber \\
& \hspace{15pt} \times\Bigl(-t_1 (-w_1+w_2+w_3+1)-t_2 (w_1-w_2+w_3+1)-t_3 (w_1+w_2-w_3+1) \nonumber \\
& \hspace{33pt} -t_2 t_1 (w_1+w_2-w_3-1)-t_3 t_1 (w_1-w_2+w_3-1)-t_2 t_3 (-w_1+w_2+w_3-1) \nonumber \\
& \hspace{33pt} +t_1 t_2 t_3 (w_1+w_2+w_3-1)+w_1+w_2+w_3+1 \Bigr)^{-j_1-j_2-j_3+\frac{k}{2}}\ , \label{eq:3pt-solution-y-basis-covering map exists}
\end{align}
where 
\be 
t_i = a_i^{-1} y_i
\ee
and $a_i$ is given in \eqref{eq:three-pt-ai}. For simplicity, we omitted the dependence on right-movers. 

Before explaining how to perform the full integral, let us first show how to compute the solution to the recursion relations for discrete representations, where the integral runs over a contour $\mathfrak{C}$ around 0 or $\infty$.  Integrating over all possible closed contours actually provides a (possibly overcomplete) basis of solutions to the recursion relations in the $h$-basis. This is analogous to the procedure that is routinely applied in the Coulomb gas formalism where one is interested in similar integrals. The chiral integrals lead to a basis of solutions that can be identified with the conformal blocks and the full integral \eqref{h-basis-3pt} is then written as a finite sum of such conformal blocks. Thus we are interested in computing the integral
\begin{align}
&C(w_1,w_2,w_3;j_1,j_2,j_3;k) \prod_{i=1}^3 a_i^{\frac{k}{4}(w_i-1)-h_i}w_i^{-\frac{k}{4}(w_i+1)+j_i} \Pi^{-\frac{k}{2}} \\
&\times  \int_\mathfrak{C} \prod_{i=1}^3  \mathrm{d}t_i\ t_i^{\frac{k w_i}{2}+j_i-h_i-1} (1-t_1)^{-j_1+j_2+j_3-\frac{k}{2}} (1-t_2)^{j_1-j_2+j_3-\frac{k}{2}} (1-t_3)^{j_1+j_2-j_3-\frac{k}{2}} \nonumber \\
&\qquad \times\Bigl(-t_1 (-w_1+w_2+w_3+1)-t_2 (w_1-w_2+w_3+1)-t_3 (w_1+w_2-w_3+1) \nonumber \\
& \qquad\qquad -t_2 t_1 (w_1+w_2-w_3-1)-t_3 t_1 (w_1-w_2+w_3-1)-t_2 t_3 (-w_1+w_2+w_3-1) \nonumber \\
& \qquad\qquad +t_1 t_2 t_3 (w_1+w_2+w_3-1)+w_1+w_2+w_3+1 \Bigr)^{-j_1-j_2-j_3+\frac{k}{2}}\ ,
\end{align}
where $\mathfrak{C}$ is the appropriate contour. We reinstated the (chiral) normalization factor and used various identities of the Barnes G function to simplify it. In this expression $N=\frac{1}{2}(w_1+w_2+w_3-1)$ and\footnote{From a geometric point of view, $\Pi$ is the product of the residues of the branched covering map ${\gamma(\zeta): \text{S}^2 \to \text{S}^2}$ with ramification indices $w_1, w_2, w_3$, i.e.~$\Pi=\prod_a \mathop{\text{Res}}_{\zeta=\zeta_a} \gamma(\zeta)$. Similarly, the $a_i$ introduced in \eqref{eq:three-pt-ai} enter the Taylor expansion of the covering map close to the covering space insertions, $\gamma(\zeta)=x_i+a_i(\zeta-z_i)^{w_i}+\mathcal{O}\big((\zeta-z_i)^{w_i+1}\big)$. We will explain the geometric picture in \cite{paper2}.}
\be 
\Pi= \frac{f(w_1)f(w_2)f(w_3)}{\Gamma\left(\frac{w_1+w_2+w_3+1}{2}\right)f\left(\frac{w_1+w_2+w_3+1}{2}\right)f\left(\frac{-w_1+w_2+w_3+1}{2}\right)f\left(\frac{w_1-w_2+w_3+1}{2}\right)f\left(\frac{w_1+w_2-w_3+1}{2}\right)} \ , 
\label{eq:definition C}
\ee
with
\be
f(n) \equiv \frac{\Gamma(n)^{n-1}}{G(n)^2}  =\prod_{k=1}^{n-1} k^{2k-n+1}\ .
\ee
Here $G$ is again the Barnes G function, see eq.~\eqref{barnesG}. One can again easily convince oneself numerically that this expression follows from the previous formulas given in \eqref{3pt-odd-parity}. The integral simplifies significantly under the substitution $t_i \to 1-t_i^{-1}$ which leads to 
\begin{multline}
\left\langle V_{j_1,h_1,\bar{h}_1}(0;0)V_{j_2,h_2,\bar{h}_2}(1;1)V_{j_3,h_3,\bar{h}_3}(\infty;\infty) \right\rangle=C(w_1,w_2,w_3;j_1,j_2,j_3;k) \\
\times N^{k-2j_1-2j_2-2j_3}\prod_{i=1}^3 a_i^{\frac{k}{2}(w_i-1)-h_i-\bar{h}_i}w_i^{-\frac{k}{2}(w_i+1)+2j_i} \Pi^{-k}  \Bigg|\int_\mathfrak{C} \prod_{i=1}^3 \mathrm{d} t_i \ t_i^{-\frac{k w_i}{2}+j_i+h_i-1} \\
\times(1-t_i)^{\frac{k w_i}{2}+j_i-h_i-1}\left(1-\frac{w_1 t_1}{N}-\frac{w_2 t_2}{N}-\frac{w_3 t_3}{N}\right)^{\frac{k}{2}-j_1-j_2-j_3}\Bigg|^2\ . \label{eq:three point function integral expression}
\end{multline}
 This integral is of hypergeometric type with linear integrands. 
\paragraph{Discrete representation $\boldsymbol{\mathcal{D}^-}$.} To continue, let us assume that we are dealing with a discrete representation which satisfies $h_i+j_i-\frac{kw_i}{2} \in \mathds{Z}$. This representation is of type $\mathcal{D}^-$.  Then we can take the contour that encircles 0 for $i=1,\,2,\,3$ (since this is where the singularity at $y_i=\infty$ that is used in the definition in Section~\ref{sec:y-transform} is mapped to under the change of variables that we performed). This simply extracts the coefficient $t_i^{-1}$ from the expression when expanded around 0. The integral clearly vanishes identically whenever $h_i+j_i-\frac{kw_i}{2} \in \mathds{Z}_{\ge 1}$. This means that the highest weight state of the representation satisfies $h_i-\frac{kw_i}{2}=-j_i$, which is indeed the representation $\mathcal{D}_{j_i}^-$.

We can now simply expand the factors in powers of $t_i$ using the binomial and multinomial theorem respectively. This leads to (dropping the prefactors momentarily)
\begin{align}
&\int_\mathfrak{C} \prod_{i=1}^3 \mathrm{d} t_i \ t_i^{-\frac{k w_i}{2}+j_i+h_i-1} \sum_{m_1,m_2,m_3,n_1,n_2,n_3=0}^\infty \binom{h_i-j_i-\frac{k w_i}{2}+m_i}{m_i} \nonumber\\
&\hspace{30pt}\times\binom{j_1+j_2+j_3-\frac{k}{2}-1+n_1+n_2+n_3}{n_1,\,n_2,\, n_3}\prod_{i=1}^3 t_i^{m_i+n_i} \left(\frac{w_i}{N} \right)^{n_i} \nonumber \\
&= \sum_{n_1,n_2,n_3=0}^\infty \binom{j_1+j_2+j_3-\frac{k}{2}-1+n_1+n_2+n_3}{n_1,\,n_2,\, n_3} \prod_{i=1}^3 \binom{-2j_i-n_i}{h_i-\frac{kw_i}{2}-j_i}  \left(\frac{w_i}{N} \right)^{n_i}\ .
\end{align}
Even though it is not obvious the infinite sums over $n_i$ actually truncate leaving us with finite sums. 

We can compare this with the definition of the Lauricella hypergeometric function of type A \cite{Lauricella}
\be 
F_A(a;b_1,b_2,b_3;c_1,c_2,c_3;x_1,x_2,x_3)\equiv\sum_{n_1,n_2,n_3=0}^\infty \frac{(a)_{n_1+n_2+n_3}\prod_i(b_i)_{n_i}}{\prod_i (c_i)_{n_i} n_i!} \prod_i x_i^{n_i}\ , \label{eq:definition Lauricella hypergeometric function}
\ee
where $(a)_n=a(a+1) \cdots(a+n-1)$ is the rising Pochhammer symbol. This power series truncates if either $b_i \in \mathds{Z}_{\le 0}$ for each $i$ or if $a \in \mathds{Z}_{\le 0}$. Otherwise, it converges only for $|x_1|+|x_2|+|x_3|<1$ but can be analytically continued to a meromorphic function on $\mathds{C}^3$ with branch loci at
\be 
\sum_{i \in I} x_i=1
\ee
for any subset $I \subset \{1,2,3\}$. It follows now that the integral we are interested in can be expressed in terms of the Lauricella hypergeometric function as follows:
\be 
\binom{j_i+\frac{kw_i}{2}-h_i-1 }{ 2j_i-1} F_A\left(j_1+j_2+j_3-\frac{k}{2};j_i+h_i-\frac{k w_i}{2};2j_i;\frac{w_i}{N} \right)\ .
\ee
As advertised, this solution does indeed vanish manifestly for $j_i+h_i-\frac{kw_i}{2} \in \mathds{Z}_{\ge 1}$. Thus, we have the three-point function for discrete representations:
\begin{align}
&\left\langle V_{j_1,h_1,\bar{h}_1}(0;y_1;0)V_{j_2,h_2,\bar{h}_2}(1;y_2;1)V_{j_3,h_3,\bar{h}_3}(\infty;y_3;\infty) \right\rangle=C(w_1,w_2,w_3;j_1,j_2,j_3;k) \nonumber\\
&\qquad\times N^{k-2j_1-2j_2-2j_3}\prod_{i=1}^3 a_i^{\frac{k}{2}(w_i-1)-h_i-\bar{h}_i}w_i^{-\frac{k}{2}(w_i+1)+2j_i} \Pi^{-k}  \nonumber\\
&\qquad\times \left|\prod_{i=1}^3\binom{j_i+\frac{kw_i}{2}-h_i-1 }{ 2j_i-1}F_A\left(j_1+j_2+j_3-\frac{k}{2};j_i+h_i-\frac{k w_i}{2};2j_i;\frac{w_i}{N} \right)\right|^2\ . 
\label{h-basis-solution-3D-}
\end{align}
As we shall discuss later, we believe that $C(w_1,w_2,w_3;j_1,j_2,j_3;k)=\mathcal N(j_1) D(\tfrac{k}{2}-j_1,j_2,j_3)$, where $D(j_1,j_2,j_3)$ is the unflowed three-point function and $\mathcal N(j)$  is defined through eq.~\eqref{Nj} in terms of the normalization of the unflowed two-point function $B(j)$, see Appendix~\ref{app:unflowed three-point functions}. Thus we have indeed given a full conjecture for the value of the three-point functions. It is clear that it is much more convenient to work in the $y$-basis for computational purposes.
\paragraph{Discrete representation $\boldsymbol{\mathcal{D}^+}$.} 
Similarly we can find the solution of three $\mathcal{D}^+_{j_i}$ by choosing the contour to run around $t_i=1$ for $i=1,\,2,\,3$. This leads to formally the same integral after the substitution, except that $h_i-\frac{kw_i}{2} \longrightarrow -h_i+\frac{kw_i}{2}$ and that $N \to N'$ where $N'=\frac{1}{2}(w_1+w_2+w_3+1)$. Including all prefactors, we get
\begin{align}
&\left\langle V_{j_1,h_1,\bar{h}_1}(0;y_1;0)V_{j_2,h_2,\bar{h}_2}(1;y_2;1)V_{j_3,h_3,\bar{h}_3}(\infty;y_3;\infty) \right\rangle=C(w_1,w_2,w_3;j_1,j_2,j_3;k) \nonumber\\
&\qquad\times (N')^{k-2j_1-2j_2-2j_3}\prod_{i=1}^3 a_i^{\frac{k}{2}(w_i-1)-h_i-\bar{h}_i}w_i^{-\frac{k}{2}(w_i+1)+2j_i} \Pi^{-k}  \nonumber\\
&\qquad\times \left|\prod_{i=1}^3 \binom{j_i-\frac{kw_i}{2}+h_i-1 }{ 2j_i-1} F_A\left(j_1+j_2+j_3-\frac{k}{2};j_i-h_i+\frac{k w_i}{2};2j_i;\frac{w_i}{N'} \right)\right|^2\ .
\label{h-basis-solution-3D+}
\end{align}
It is clear that we could repeat this calculation with any mixture of $\mathcal{D}^+$ and $\mathcal{D}^-$ representations.

\paragraph{Continuous representation.} This is the most complicated of the cases since we should compute the full non-holomorphic integral over the $y$-variable. This can be done by decomposing the integral into a finite sum of holomorphic times antiholomorphic integrals, similar to the KLT formula \cite{Kawai:1985xq} that relates closed string integrals to open string integrals. We compute the integral over the $y$-space using twisted cohomology. Since this is quite technical, we relegated it to Appendix~\ref{app:integral Lauricella}. The result is expressed as a sum of eight Lauricella hypergeometric functions, see eq.~\eqref{h-basis-3pt-C}.

\subsection{Fusion rules}
\label{subsec:fusion rules}

As a byproduct of our analysis we can read off the fusion rules of the theory (or rather the non-vanishing three-point functions from which we will deduce the fusion rules). 
Let us analyze the case of even parity to illustrate how these fusion rules arise. We gave the corresponding solution in the $y$-basis representation in \eqref{3pt-even-parity}. We can clearly find non-vanishing correlators with continuous representation of this type by integrating the correlator over all of $y$-space, similarly as we did in Appendix~\ref{app:integral Lauricella} (but the resulting hypergeometric function is of a different kind). The question whether there are non-vanishing three-point functions with discrete representations now boils down to establishing whether the contour integrals around the corresponding points $y_i=0$ or $y_i=\infty$ are non-vanishing.

For illustration, let us consider the correlator with three $\mathcal{D}^-_j$ representations. To obtain it, we expand \eqref{3pt-even-parity} around $(y_1,y_2,y_3)=(\infty,\infty,\infty)$, as suggested by eq.~\eqref{eq:y-inverse-D-}. This leads to
\be 
P_{\boldsymbol{w}-e_{12}}P_{\boldsymbol{w}-e_{13}}P_{\boldsymbol{w}-e_{23}}\int _\infty \mathrm{d}y_1\ \mathrm{d}y_2\ \mathrm{d}y_3\  \prod_{i=1}^3 y_i^{\frac{kw_i}{2}-j_i-h_i-1} \left(1+\mathcal{O}\left(y_1^{-1},y_2^{-1},y_3^{-1}\right)\right)\ .
\ee
This has indeed the correct dependence on $y_i$ to lead to three $\mathcal{D}_j^-$ representations. However, we have to make sure that the prefactor is non-zero. It follows directly from the definition in Section~\ref{sec:y-dependence-3ptf} that vanishing of the prefactor is equivalent to
\be 
\sum_i w_i =2\max_i w_i\ .
\ee
Thus in this case, some fusion channels are forbidden. One can similarly analyze three-point functions with mixed representations. For the three point function of the form $\langle [\mathcal{D}^{\varepsilon_1}]^{w_1}[\mathcal{D}^{\varepsilon_2}]^{w_2}[\mathcal{D}^{\varepsilon_3}]^{w_3}\rangle$ with $\varepsilon_i=\pm 1$, the condition for the three-point function to be non-vanishing takes the form
\be 
\prod_{1 \le i<j\le 3} P_{\boldsymbol{w}+\varepsilon_i e_i+\varepsilon_j e_j}= 0\ .
\ee
Overall, this analysis (together with the odd parity case) produces selection rules that can be summarized as follows:
\begin{enumerate}
\item $\langle [\mathcal{D}^{\varepsilon_1}]^{w_1}[\mathcal{D}^{\varepsilon_2}]^{w_2}[\mathcal{D}^{\varepsilon_3}]^{w_3} \rangle \ne 0$ $\Leftrightarrow$ $\sum_i \left(w_i+\frac{\varepsilon_i-1}{2}\right) \ge 2\max_i \left(w_i+\frac{\varepsilon_i-1}{2}\right)$ for $i=1,\,2,\,3$. Here $\varepsilon_i =\pm 1$, determines the type of discrete representation.
\item $\langle [\mathcal{R}_1]^{w_1}[\mathcal{R}_2]^{w_2} [\mathcal{C}]^{w_3} \rangle \ne 0$ $\Leftrightarrow$ $\langle [ \mathcal{R}_1]^{w_1}[ \mathcal{R}_2 ]^{w_2}[\mathcal{D}^+]^{w_3}\rangle \ne 0$ or $\langle [ \mathcal{R}_1]^{w_1}[ \mathcal{R}_2 ]^{w_2}[\mathcal{D}^-]^{w_3}\rangle \ne 0$. Here $\mathcal{R}_1$ and $\mathcal{R}_2$ are any of $\mathcal{D}^+$, $\mathcal{D}^-$ or $\mathcal{C}$.
\end{enumerate}
These rules are very intuitive. Since $[\mathcal{D}^+_{j}]^w \cong [\mathcal{D}^-_{\frac{k}{2}-j}]^{w+1}$, one would expect that only the quantity $w_i+\frac{\varepsilon_i-1}{2}$ can enter the fusion rules for discrete representations.\footnote{This isomorphism is valid as a representation. We are nonetheless considering different states in the correlators. This can be seen in the right Figure~\ref{fig:Verma module}. The states on the orange strip correspond to the affine primary states w.r.t.~$[\mathcal{D}^-]^{w+1}$ and the states on the blue strip to the affine primary states w.r.t.~$[\mathcal{D}^+]^w$. There is exactly one overlap state for which we could make contact using different representations of the state either inside a $[\mathcal{D}^+]^w$ or a $[\mathcal{D}^-]^{w+1}$ representation.}
The second rule is also sensible since the three types of representations fit in a short exact sequence
\be 
0 \longrightarrow [\mathcal{D}^+_j]^w \longrightarrow [\mathcal{C}_j ]^w\longrightarrow [\mathcal{D}^-_{1-j}]^w \longrightarrow 0\ .
\ee
Hence on the level of states (more precisely the Grothendieck ring), $[\mathcal{C}_j ]^w\sim [\mathcal{D}^+_j ]^w \oplus [\mathcal{D}^-_{1-j}]^w $, thus leading to the second rule. In these rules, there is no restriction on $j$, i.e.\ the three-point functions can be non-zero for any choice of spins $j_1$, $j_2$ and $j_3$ in the allowed ranges discussed in section~\ref{sec:spectral flow}. The exception to this statement is given by spins $j$ corresponding to degenerate spins $j$.  These constitute a measure zero of all possible spins and we focus here on the `generic' three-point functions.

We can restate these rules also in terms of fusion. For this, we first observe that the two-point function can be viewed as a non-degenerate metric on the space of degenerate representations. The following two-point functions are non-vanishing:
\be 
\langle [\mathcal{D}^+_j]^w [\mathcal{D}^+_j]^w \rangle\ , \qquad \langle [\mathcal{D}^-_j]^w [\mathcal{D}^-_j]^w \rangle\ , \qquad \langle [\mathcal{C}_j]^w [\mathcal{C}_j]^w \rangle\ , \qquad 
\langle [\mathcal{C}_j]^w [\mathcal{C}_{1-j}]^w \rangle\ , 
\ee
so that every representation is self-dual. Thus the existence of the fusion rule $[\mathcal{D}^+]^{w_1} \times [\mathcal{D}^+]^{w_2} \longrightarrow [\mathcal{D}^+]^{w_3}$ 
is equivalent to the non-vanishing of $\langle [\mathcal{D}^+]^{w_1} [\mathcal{D}^+]^{w_2} [\mathcal{D}^+]^{w_3}\rangle $ and similarly for other representations. There will be again exceptional fusion rules for degenerate representations, where additionally the spins that appear on the right-hand side of the fusion rule are restricted.

\subsection{Spectral flow violation}
\label{sec:spectral flow violation}

This subsection serves mostly as a guide to how our result is connected to previous literature \cite{Fateev,Maldacena:2000kv,Giribet:2000fy,Giribet:2001ft,Giribet:2005ix,Ribault:2005ms,Giribet:2005mc,Minces:2005nb,Iguri:2007af,Baron:2008qf,Iguri:2009cf,Giribet:2011xf,Cagnacci:2013ufa,Giribet:2015oiy,Giribet:2019new,Hikida:2020kil}. The main innovation is a much better control over the positions of the vertex operators in the dual CFT space $x$. In previous literature often the special case of vertex operators with $x=0$ or $x=\infty$ is considered. This simplifies the analysis dramatically. While we have mostly considered the case $(x_1=0,x_2=1,x_3=\infty)$, we can use the solution of the global Ward identities \eqref{eq:global Ward identities solution 3pt function} to reconstruct the solution for three generic points $x_i$. In particular, we can analyze under what circumstances we can set $(x_1,x_2,x_3)=(0,x,\infty)$, take the limit $x \to 0$ and still get a well-defined result. For simplicity, let us assume in the following that $w_i\ge 1$ for all $i$ so that some pathological cases can be excluded. We have
\begin{multline}
\left \langle V_{j_1}(0;y_1;0) V_{j_2}(x;y_2;1)V_{j_3}(\infty;y_3;\infty)\right\rangle \\
=x^{-j_1-j_2+j_3+\frac{k}{2}(-w_1-w_2+w_3)} \left \langle V_{j_1}\left(0;\frac{y_1}{x};0\right) V_{j_2}\left(1;\frac{y_2}{x};1\right)V_{j_3}(\infty;y_3 x;\infty)\right\rangle
\end{multline}
Now in the limit $x \to 0$, certain terms in the definition of $X_I$ dominate. There are three cases in which we get a regular limit $x \to 0$. They occur for
\be 
|w_1+w_2-w_3| \le 1\ . \label{eq:spectral flow violation}
\ee
Thus in these cases there is a well-defined three-point function with two coincident insertion points in $x$-space. In these cases one can of course recover the three-point function at three general points $(x_1,x_2,x_3)$ because it follows from the global Ward identities. It is this correlation function that was usually studied previously in the literature. The bound \eqref{eq:spectral flow violation} was referred to as `spectral flow violation', i.e.\ spectral flow in a three-point function is conserved up to possibly one unit \cite{Fateev,Maldacena:2001km}. 

The three-point functions for $(x_1=0,x_2=0,x_3=\infty)$ are moreover quite easy to state.  They only depend on $w_1+w_2-w_3$, up to signs. They take the explicit form (here we assume that $w_i \ge 1$ for all $i$, so that all three $y_i$'s are present)
\begin{subequations}
\begin{align}
& w_1+w_2-w_3=1:  \nonumber \\
& \hspace{1.5cm} \left\langle V_{j_1}^{w_1}(0;(-1)^{w_1} y_1;0)V_{j_2}^{w_2}(0;(-1)^{w_3} y_2;1)V_{j_3}^{w_3}(\infty; y_3;\infty)\right\rangle\nonumber\\
& \hspace{3cm} =y_1^{-j_1+j_2+j_3-\frac{k}{2}}y_2^{j_1-j_2+j_3-\frac{k}{2}}\left(y_1+y_2+y_1y_2y_3\right)^{\frac{k}{2}-j_1-j_2-j_3}\ , \\
& w_1+w_2-w_3=0: \nonumber \\
& \hspace{1.5cm}\left\langle V_{j_1}^{w_1}(0;(-1)^{w_1} y_1;0)V_{j_2}^{w_2}(0;(-1)^{w_3} y_2;1)V_{j_3}^{w_3}(\infty; y_3;\infty)\right\rangle\nonumber\\
& \hspace{3cm} =(y_1-y_2)^{-j_1-j_2+j_3}(1+y_1y_3)^{-j_1+j_2-j_3}\left(1+y_2y_3\right)^{j_1-j_2-j_3}\ , \\
& w_1+w_2-w_3=-1: \nonumber \\
& \hspace{1.5cm} \left\langle V_{j_1}^{w_1}(0;(-1)^{w_1} y_1;0)V_{j_2}^{w_2}(0;(-1)^{w_3} y_2;1)V_{j_3}^{w_3}(\infty; y_3;\infty)\right\rangle\nonumber\\
& \hspace{3cm} =y_3^{j_1+j_2-j_3-\frac{k}{2}}\left(1+y_1y_3+y_2y_3\right)^{\frac{k}{2}-j_1-j_2-j_3}\ .
\end{align}
\end{subequations}
We also suppressed normalization constants of the answer. One could of course compute again an integral transform to express this in the $h$-basis. In the other cases, for $w_1+w_2-w_3=\delta\ge 2$, we find that the three-point function diverges as $\mathcal{O}(x^{j_1+j_2+j_3-\frac{k \delta}{2}})$ as $x \to 0$.
\medskip

We think that our investigation has clarified the origin of the spectral flow violation rule \eqref{eq:spectral flow violation}. In particular,  this leads to more restrictive fusion rules than the ones we found in Section~\ref{subsec:fusion rules}. This is however an artifact of the too restrictive choice of states in the Verma module. What we have shown is that if one looks at coherent states obtained by acting with $\mathrm{e}^{x J_0^+}$ in the Verma module then more fusion channels are possible. 
It is of course a logical possibility that we also still miss fusion channels because one should look at even more general states in the Verma module than the ones we did. We believe this not to be the case.

We should hasten to say that the bounds \eqref{eq:spectral flow violation} have an important significance in holography and hence the study of these restricted fusion channels is well-motivated. In fact, there is an important class of operators with regular OPEs in the dual CFT (i.e.\ $x$-space) in the $\mathcal{N}=(2,2)$ supersymmetric setup. These are chiral primaries forming the chiral ring and the corresponding three-point functions are extremal three-point functions. Hence to study them it is sufficient to consider vertex operators at $x=0$ or $x=\infty$.

\subsection{Examples and comparison with previous results}
\label{sec:3pt examples}
In order to clarify our construction, let us discuss some examples and compare with previous results in the literature.  In this section, we only consider continuous representations.

\paragraph{$\boldsymbol{(w_1, w_2, w_3)=(1,0,0)}$} Let us start from the simplest example, the correlator
\begin{multline}
\left\langle V^{1}_{j_1, h_1}(0;0) \,  V^{0}_{j_2}(1;1) \, V^{0}_{j_3}(\infty; \infty) \right\rangle \\
= \, C(1,0,0; j_i; k)  \int \mathrm{d}^2 y_1 \, y_1^{\frac{k}{2}+j_1-h_1-1} \bar y_1^{\frac{k}{2}+j_1-\bar h_1-1} \,  |X_{123}|^{2(\frac{k}{2}-j_1-j_2-j_3)} \\
\times \  |X_1|^{2(-\frac{k}{2}-j_1+j_2+j_3)} |X_2|^{2(-\frac{k}{2}+j_1-j_2+j_3)} |X_3|^{2(-\frac{k}{2}+j_1+j_2-j_3)} \ , 
\label{100-example}
\end{multline}
where we have made use of eqs.~\eqref{3pt-odd-parity} and \eqref{h-basis-3pt}. One can easily check that (see \eqref{X_I-3pt} and \eqref{Pw-definition})
\be
X_{123} = -1 + y_1 \ , \quad X_1= y_1 \ , \quad X_2 = 1 \ , \quad X_3 = 1 \ . 
\ee
The integral in \eqref{100-example} can then be performed explicitly --- see eq.~\eqref{eq:three-singular-points} --- and it reads 
\begin{align}
& \left\langle V^{1}_{j_1, h_1}(0;0) \,  V^{0}_{j_2}(1;1) \, V^{0}_{j_3}(\infty; \infty) \right\rangle \nonumber \\ 
& \qquad = \pi C(1,0,0; j_i; k)\frac{\lgamma(j_2+j_3-h_1)\lgamma(\frac{k}{2}-j_1-j_2-j_3+1)}{\lgamma(\frac{k}{2}-j_1-h_1+1)} \ . 
\label{100-example-result}
\end{align}
Eq.~\eqref{100-example-result} reproduces exactly what Maldacena and Ooguri computed in \cite{Maldacena:2001km} by different techniques, with 
\be 
C\bigl(1,0,0; j_i; k \bigr) =  \, \mathcal N(j_1) \, D\left(\tfrac{k}{2}-j_1, j_2, j_3\right) \ ,
\ee
and 
\be
\mathcal N(j) \equiv \sqrt{\frac{B(j)}{B(\frac{k}{2}-j)}} \ ,
\label{Nj} 
\ee
see their eq.~(E.18) and make the identifications $h_1 = J^{\text{MO}}$, $\bar h_1 = \bar J^{\text{MO}}$, $j_1 = j_1^{\text{MO}}$, $j_2=j_3^{\text{MO}}$, $j_3=j_4^{\text{MO}}$.\footnote{There is a $k$-dependent undetermined constant in eq.~(E.18) of \cite{Maldacena:2001km}. It can be recovered explicitly by tracing it back through the various steps of their computation.} See Appendix \ref{app:unflowed three-point functions} for the precise form of $B(j)$ and of the unflowed structure constants $D(j_1, j_2, j_3)$.

\paragraph{$\boldsymbol{(w_1, w_2, w_3)=(w,w,0)}$} In this case the correlator reads
\begin{multline}
\left\langle V^{w}_{j_1}(0;0) \,  V^{w}_{j_2}(1; 1) \, V^{0}_{j_3}(\infty; \infty) \right\rangle \\
= C\bigl(w,w,0; j_i; k \bigr) \int \mathrm{d}^2 y_1\, \mathrm{d}^2y_2 \, \prod_{i=1,2} y_i^{\frac{kw}{2}+j_i-h_i-1} \bar y_i^{\frac{kw}{2}+j_i-\bar h_i-1} \\
\times \, |1-y_1|^{2(-j_1+j_2-j_3)}  |1-y_2|^{2(j_1-j_2-j_3)} |1-y_1y_2|^{2(-j_1-j_2+j_3)} \ , 
\label{ww0-example}
\end{multline}
which upon setting 
\be 
C\bigl(w,w,0; j_i; k \bigr) =  \, D\left(j_1, j_2, j_3\right) \ ,
\ee
 agrees with the result of \cite{Cagnacci:2013ufa}, see their eqs.~(62), (80) and (81) and identify $h_1 = J_1^{\text{CI}}$, $h_2 = J_2^{\text{CI}}$, $j_1 = j_1^{\text{CI}}$, $j_2 = j_2^{\text{CI}}$, $j_3 = j_3^{\text{CI}}$. Proceeding along the lines of \cite{Teschner:1999ug,Maldacena:2001km,Cagnacci:2013ufa} one can show that \eqref{ww0-example} correctly reproduces the two-point function \eqref{eq:2ptf-final} for $j_3=0$ (including the normalization $D(j_1,j_2,j_3)$). 

\paragraph{$\boldsymbol{(w_1, w_2, w_3)=(w+1,w,0)}$} Similarly to the previous example, we have
\begin{align}
& \left\langle V^{w+1}_{j_1}(0; 0) \,  V^{w}_{j_2}(1; 1) \, V^{0}_{j_3}(\infty; \infty) \right\rangle \nonumber \\
& \quad = C\bigl(w+1,w,0; j_i; k \bigr) \int \mathrm{d}^2 y_1\,  \mathrm{d}^2y_2 \, y_1^{\frac{kw}{2}-h_1+j_2+j_3-1} \bar y_1^{\frac{kw}{2}-\bar h_1+j_2+j_3-1}  \nonumber \\ 
& \hspace{120pt}  \times \, y_2^{\frac{kw}{2}+j_2-h_2-1} \bar y_2^{\frac{kw}{2}+j_2-\bar h_2-1} |1+y_1-y_1y_2|^{2(\frac{k}{2}-j_1-j_2-j_3)} \\
& \quad = C\bigl(w+1,w,0; j_i; k \bigr) \int \mathrm{d}^2 y_1\,  \mathrm{d}^2y_2 \, y_1^{-\frac{k(w+1)}{2}+j_1+h_1-1} \bar y_1^{-\frac{k(w+1)}{2}+j_1+\bar h_1-1}  \nonumber \\
& \hspace{120pt} \times \, y_2^{\frac{kw}{2}+j_2-h_2-1} \bar y_2^{\frac{kw}{2}+j_2-\bar h_2-1} |1-y_1-y_2|^{2(\frac{k}{2}-j_1-j_2-j_3)} \label{wp1w0-example1} \\
& \quad = \pi^2C\bigl(w+1,w,0; j_i; k \bigr)   \frac{ \lgamma\bigl(\frac{k}{2}-j_1-j_2-j_3+1\bigr)\lgamma\bigl(h_1-\frac{k(w+1)}{2}+j_1\bigr) \lgamma\bigl(\frac{kw}{2}+j_2-h_2\bigr) }{\lgamma(h_1-h_2-j_3+1)} \ , 
\label{wp1w0-example}
\end{align}
where in \eqref{wp1w0-example1} we changed the integration variable $y_1 \to y_1^{-1}$, while in \eqref{wp1w0-example} we used the identity \eqref{general-integral}. Upon identifying $j_1 = j_1^{\text{CI}}$, $j_2 = j_2^{\text{CI}}$, $h_1= J_1^{\text{CI}}$, $h_2 = J_2^{\text{CI}}$, $j_3 = J_3^{\text{CI}}$, eq.~\eqref{wp1w0-example} exactly reproduces the results of \cite{Cagnacci:2013ufa}, with\footnote{In deriving eq.~\eqref{eq:normalisation-w-wp1-0} one has to take into account various $k$-dependent factors that are usually neglected in the literature.} 
\be 
C\bigl(w+1,w,0; j_i; k \bigr) =  \, \mathcal N(j_1) \, D\left(\tfrac{k}{2}-j_1, j_2, j_3\right) 
\label{eq:normalisation-w-wp1-0}
\ee
and $\mathcal N(j)$ is defined in eq.~\eqref{Nj}.

\subsection{A conjecture for the normalisation}
\label{sec:normalisation-conjecture}

We have seen in the examples of the previous section that the normalization that was unfixed by symmetry simply took the form
\be 
C(w_1,w_2,w_3;j_1,j_2,j_3;k)=\begin{cases}
D(j_1,j_2,j_3)\ , \quad w_1+w_2+w_3 \in 2\mathds{Z}\ , \\
\mathcal N(j_1)D(\tfrac{k}{2}-j_1,j_2,j_3)\ , \quad w_1+w_2+w_3 \in 2\mathds{Z}+1\ .
\end{cases}
\ee
The second case looks asymmetric in the three variables, but this is actually not the case thanks to identity \eqref{eq:symmetry shifted three point function}. In fact, identities \eqref{eq:symmetry shifted three point function} and \eqref{eq:symmetry shifted three point function 2} imply that we could have replaced any number of spins $j_i \to \frac{k}{2}-j_i$, as long as we swap an even number for the even parity case and an odd number of the odd parity case. This makes our conjecture very natural and in fact this completely fixes the normalization constant $\mathcal{N}(j)$.
We should mention that we put an explicit normalization factor of $X_\emptyset^{j_1+j_2+j_3-k}$ in the even parity solution \eqref{3pt-even-parity}. For the case $\boldsymbol{w}=(w,w,0)$ that we have compared to the example, this factor is 1 and hence not observable. We nonetheless think that its presence is natural, because in this way both the even and odd parity cases have four factors. More convincingly, this factor is present together with a non-trivial function of the crossratios in the four-point function to which we return in a future publication \cite{paper2}.
We conjecture that these formulae hold true in general, regardless of the values of $(w_1,w_2,w_3)$. Below we collect some further  evidence for this by showing that this factor is necessary for the result to be reflection symmetric. More evidence will follow in a future publication \cite{paper2}.

To summarize, we conjecture that the three-point functions take the following form
\begin{tcolorbox}%
\begin{align}
&\left\langle V^{w_1}_{j_1}(0; y_1; 0) \,  V^{w_2}_{j_2}(1; y_2; 1) \, V^{w_3}_{j_3}(\infty; y_3; \infty) \right\rangle \nonumber\\
&\quad=\begin{cases} D(j_1,j_2,j_3)|X_\emptyset|^{2\sum_l j_l-2k}\displaystyle\prod_{i<\ell}^3  |X_{i \ell}|^{2\sum_l j_l-4j_i-4j_\ell} \  , & \sum_i w_i \in 2\mathds{Z}\ ,\\
\mathcal N(j_1)D(\tfrac{k}{2}-j_1,j_2,j_3) |X_{123}|^{k-2\sum_l j_l} \displaystyle\prod_{i=1}^3 |X_i|^{2\sum_l j_l-4j_i-k} \ ,  & \sum_i w_i \in 2\mathds{Z}+1\ .
\end{cases} \label{eq:main conjecture}
\end{align}
\end{tcolorbox}%
This is the main result of our paper. Let us point the reader to the various definitions that enter this formula. The quantities $X_I$ that are defined for any subset $I \subset \{1,2,3\}$ are given in \eqref{X_I-3pt}. $D(j_1,j_2,j_3)$ are the unflowed three-point functions that are given in eq.~\eqref{eq:unflowed three point function}. $\mathcal N(j)$ is defined in eq.~\eqref{Nj} in terms of $B(j)$, which is the normalization of the unflowed two-point function \eqref{eq:normalization vertex operators}, see eq.~\eqref{eq:B}.

\subsection{Reflection symmetry}
We subject our proposal \eqref{eq:main conjecture} to one further consistency check. If we consider a continuous representation, then it is a non-trivial check that our formula is reflection symmetric. We now show that this is the case.

\paragraph{Parity even case.} Let us discuss the reflection symmetry first in the parity-even case (i.e.~$\sum_i w_i \in 2\mathds{Z}$).  To demonstrate reflection symmetry, we work directly in the $y$-basis, where reflection symmetry acts according to \eqref{eq:y-basis-reflection}. Let us reflect $j_3$, the argument for the other spins is similar. 

So we want to compute the integral
\be 
\int \mathrm{d}^2 y_3 \ |y-y_3|^{4j_3-4} |X_{13}|^{2j_1-2j_2-2j_3}|X_{23}|^{-2j_1+2j_2-2j_3} \ ,
\ee
where we omitted all factors in \eqref{eq:main conjecture} that do not depend on $y_3$. One can explicitly compute this integral by changing variables to 
\be 
u=\frac{(A_2 \, y+B_2)(A_1\, y_3+B_1)}{(A_1\, y_1+B_1)(A_2 \, y_3+B_2)} \ ,
\ee
where we have temporarily written
\be 
X_{13}=A_1 \, y_3+B_1\ , \qquad X_{23}=A_2 \, y_3+B_2\ .
\ee
This change of variables reduces the integral to
\begin{multline} 
|A_1 B_2-B_1 A_2|^{2-4j_3} |A_1 \,y+B_1 |^{2j_1-2j_2+2j_3-2}|A_2 \, y+B_2|^{-2j_1+2j_2+2j_3-2}  \\
\times\int \mathrm{d}^2 u |u|^{2j_1-2j_2-2j_3} |1-u|^{4j_3-4}\ . 
\end{multline}
We have $A_i \, y+B_i=X_{i3}(y)$, by which we mean $X_{i3}$ where $y_3$ has been replaced by $y$. Furthermore, we have the following identity:
\be 
A_1 B_2-B_1 A_2=\pm X_\emptyset X_{12}\ .
\ee
We have not tried to give a general proof of this identity, but it is easy to convince oneself numerically of its validity. We thus compute
\begin{align}
&R_{1-j_3}\frac{1-2j_3}{\pi}\int \mathrm{d}^2 y_3  \ |y-y_3|^{4j_3-4} \langle V_{j_1}^{w_1}(0;y_1;0)V_{j_2}^{w_2}(1;y_2;1)V_{j_3}^{w_3}(\infty;y_3;\infty)\rangle \\
&\qquad=R_{1-j_3}\frac{(1-2j_3)\lgamma(j_1-j_2-j_3+1)\lgamma(2j_3-1)}{\lgamma(j_1-j_2+j_3)} D(j_1,j_2,j_3)|X_\emptyset|^{2j_1+2j_2+2(1-j_3)}\nonumber\\
&\qquad\qquad\times |X_{12}|^{-2j_1-2j_2+2(1-j_3)}|X_{13}(y)|^{2j_1-2j_2-2(1-j_3)}|X_{23}(y)|^{-2j_1+2j_2-2(1-j_3)}\ .
\end{align}
This agrees with the same three-point function with $j_3 \to 1-j_3$, provided that
\be 
R_{1-j_3}\frac{(1-2j_3)\lgamma(j_1-j_2-j_3+1)\lgamma(2j_3-1)}{\lgamma(j_1-j_2+j_3)} D(j_1,j_2,j_3)=D(j_1,j_2,1-j_3)\ .
\ee
This is the statement of reflection symmetry in the unflowed sector and is eq.~\eqref{eq:reflection symmetry unflowed three point function}. Hence reflection symmetry in the flowed sector is a consequence of reflection symmetry in the unflowed sector.

\paragraph{Parity odd case.} For the spins $j_2$ and $j_3$, the argument is almost the same as above and reduces reflection symmetry in the flowed sector to reflection symmetry in the unflowed sector. This is also the case for the spin $j_1$, provided we write
\be 
\mathcal N(j_2) D(j_1,\tfrac{k}{2}-j_2,j_3)=\mathcal N(j_1) D(\tfrac{k}{2}-j_1,j_2,j_3)
\ee
for the prefactor, in which case, the statement reduces to reflection symmetry of $D(j_1,\tfrac{k}{2}-j_2,j_3)$.

\section{Conclusions, discussion and open questions}
\label{sec:conclusions}

In the following, we discuss various questions that we feel are not understood or deserve a better explanation. We keep this list short, since we will continue our exploration in \cite{paper2}.

\paragraph{Path integral proof.} There is an alternative route to obtain results in the worldsheet theory, namely via the Wakimoto representation. This is the analogue of the Coulomb gas representation of Liouville theory. This formulation has the advantage that it possesses a free-field action with a particular screening charge operator. This representation was used successfully to derive the $H_3^+$/Liouville correspondence (for unflowed correlators) and to compute some correlators in the $m$-basis \cite{Giribet:2000fy,Giribet:2001ft,Iguri:2007af,Hikida:2007tq,Giribet:2015oiy}. In light of this, it seems conceivable to give a direct path-integral derivation of our proposed formula. 

\paragraph{Supersymmetric and extremal correlators.} It is straightforward to extend our analysis to the supersymmetric case. Superstrings on $\mathrm{AdS}_3$ are described in the RNS formalism by the $\mathrm{SL}(2,\mathds{R})_{k+2}$ model and three free fermions. The shift in the level and the contributions of the free fermions can be readily incorporated in our formulas. In the supersymmetric context, it is particularly interesting to consider extremal correlators. They are protected against marginal deformations \cite{deBoer:2008ss, Baggio:2012rr} and are hence useful as probe to test the conjectured duality between superstring theory on $\mathrm{AdS}_3 \times \mathrm{S}^3 \times \mathbb{T}^4$ and the symmetric orbifold $\text{Sym}^N(\mathbb{T}^4)$ (or rather a deformation thereof in the case of higher NS-NS flux). It was confirmed in \cite{Gaberdiel:2007vu, Dabholkar:2007ey, Pakman:2007hn} that the unflowed extremal three-point functions match with those of the symmetric orbifold. This was later also extended to extremal four-point functions \cite{Cardona:2010qf}.\footnote{Somewhat surprisingly also a perfect match with the boundary correlator was found for certain non-extremal correlators of chiral primaries.} Extremal correlators precisely satisfy the relation $\sum_i j_i=\frac{k(n-2)}{2}-(n-3)$ where the correlators universal behaviour is described by the Liouville answer (and is hence simple). Spectrally flowed extremal three-point functions were also matched with the boundary CFT in \cite{Giribet:2007wp, Cardona:2009hk}. These correlators are much better explored than the ones away from extremality that we have considered in this paper. Thus it would be interesting to compare our results with these computations.

\paragraph{The spectral flow operator.} The traditional way to compute correlators with spectral flow is via the spectral flow operator $\Phi$. $\Phi$ is a special unflowed vertex operator with spin $j=\frac{k}{2}$ that possesses a null descendant. Similar to picture changing operators in superstring theory, one can then insert a number $\sum_i w_i$ of these spectral flow operators in an unflowed correlator. Taking the collision limit of these spectral flow operators with the unflowed vertex operators yields the flowed correlator \cite{Fateev,Maldacena:2001km}. This method of computation is rather cumbersome because the number of necessary spectral flow operators increases rapidly with the spectral flow which is why only correlators for very low amounts of spectral flow have been computed in this fashion. In section~\ref{sec:3pt examples}, we compared our correlators to some that were computed using the spectral flow operator in \cite{Maldacena:2001km, Minces:2005nb, Fukuda:2001jd, Satoh:2001bi, Cagnacci:2013ufa}.
Nevertheless, we were not able to make this connection direct and we feel that it would be important to do so in order to provide a cross check between the different methods. It seems conceivable to give a recursive proof in the spectral flow of our conjecture by using the spectral flow operator.

\paragraph{Relation to integrability.} Recently, the hexagon tessellation approach to correlation functions, originally developed for $\text{AdS}_5 \times \text{S}^5$ \cite{Basso:2015zoa, Eden:2016xvg, Fleury:2016ykk}, has been extended to $\text{AdS}_3 \times \text{S}^3 \times \mathbb{T}^4$ \cite{Eden:2021xhe} and some \emph{extremal} correlators \cite{Gaberdiel:2007vu, Dabholkar:2007ey, Pakman:2007hn} have been reproduced by integrability techniques. \emph{Non-extremal} correlators feature a much richer structure, in particular when spectral flow is present. In fact, from the point of view of the RNS formalism, it is clear that  correlators with insertions of spectrally flowed states will be expressed in terms of the generalized hypergeometric functions we encountered in Section \ref{sec:3ptf}. It would then be very interesting to see how this rich functional dependence emerges in the integrability approach to correlation functions.

\acknowledgments We thank Hanno Bertle, Matthias Gaberdiel, Bob Knighton, Juan Maldacena, Sebastian Mizera, Hirosi Ooguri, Sylvain Ribault, Volker Schomerus, Alessandro Sfondrini, Joerg Teschner, Cumrun Vafa and Xi Yin for interesting and stimulating discussions. We are grateful to Hanno Bertle, Matthias Gaberdiel, Rajesh Gopakumar and Bob Knighton for collaboration on related projects and to Sylvain Ribault and Alessandro Sfondrini for their comments on a preliminary version of this paper. The work of AD is funded by the Swiss National Science Foundation via the Early Postdoc.Mobility fellowship.  LE is supported by the IBM Einstein Fellowship at the Institute for Advanced Study. 	

\appendix
\section{Special functions} \label{app:special functions}
In this appendix, we fix our conventions for special functions and record useful properties.
\subsection{Barnes G-function}
The Barnes G-function is defined for positive integers as
\be 
G(n)=\prod_{i=1}^{n-1} \Gamma(i)\ ,
\ee
and $G(n)=0$ for $n \in \mathds{Z}_{\le 0}$. It can be extended to an entire function on the complex plane satisfying the functional identity
\be 
G(z+1)=\Gamma(z)G(z)\ .
\ee
We shall not need the numerous other interesting properties this function satisfies.
\subsection{Barnes zeta function}
We need this function only to define the double Gamma-function below. We set
\be 
\zeta_2(s,z\, |\, a_1,a_2)=\sum_{n_1,n_2\ge 0} (z+n_1 a_1+n_2 a_2)^{-s}\ ,
\ee
which converges to a holomorphic function for $\Re(s) \ge 2$. This definition has an obvious generalization where we sum over $N$ integers in the definition, but we shall not have a use for it. The corresponding definition $\zeta_1(s,z|a)$ is basically the Hurwitz zeta function.
It admits a meromorphic analytic continuation in the $s$-plane with simple poles at $s=1$ and $s=2$. Without loss of generality one can set $a_1=1$, since this can be absorbed in a rescaling of $z$.

\subsection{Double gamma function}

The double gamma function is defined as
\be 
\log \Gamma_2(z\, |\, 1,\omega)=\frac{\mathrm{d}}{\mathrm{d}s}\Big|_{s=0} \zeta_2(s,z\, |\, 1,\omega)\ .
\ee
Notice that this definition requires the analytic continuation of $\zeta_2(s,z|1,\omega)$ outside of the domain $\Re(s) > 2$ where it was initially defined. A computationally more useful formula is in terms of the Weierstrass form
\be 
\Gamma_2(z\, |\, 1,\omega)=\frac{\mathrm{e}^{\lambda_1 z+\lambda_2 z^2}}{z} \prod_{\begin{subarray}{c} n_1 \ge 0,\, n_2 \ge 0 \\ (n_1,n_2) \ne (0,0) \end{subarray}} f\left(\frac{z}{n_1+n_2 \omega} \right)\ ,
\ee
where
\be 
f(x)=\frac{\mathrm{e}^{x-\frac{1}{2}x^2}}{1+x}\ ,
\ee
and the coefficients $\lambda_1$ and $\lambda_2$ are defined in terms of the Laurent series expansion of $\zeta(s,z\, | \, 1,\omega)$ at $s=1$ and $s=2$,
\begin{subequations}
\begin{align}
\lambda_1&=[(s-1)^0] \, \zeta(s,0\, | \, 1,\omega)\ , \\
\lambda_1&=\frac{1}{2}[(s-2)^0] \, \zeta(s,0\, | \, 1,\omega)+\frac{1}{2}[(s-2)^{-1}] \, \zeta(s,0\, | \, 1,\omega)\ ,
\end{align}
\end{subequations}
where $[(s-s_0)^n]$ denotes the corresponding Laurent series coefficient.
The double Gamma function satisfies the following fundamental functional identities
\begin{subequations}
\begin{align}
\Gamma_2(z\, |\, 1, \omega)&=\frac{1}{\sqrt{2\pi}} \Gamma(z) \Gamma_2(z+\omega\, |\, 1,\omega)\ , \\
\Gamma_2(z\, | \, 1,\omega)&=\frac{\omega^{\frac{z}{\omega}-\frac{1}{2}}}{\sqrt{2\pi}} \Gamma\left(\frac{z}{\omega}\right)\Gamma_2(z+1\, | \, 1,\omega)\ .
\end{align}
\end{subequations}
The double Gamma function is a meromorphic function in the $z$-plane with poles at
\be 
z=-m-n \omega\, \quad m,n \in \mathds{Z}_{\ge 0}\ .
\ee
Provided that $1$ and $\omega$ are linearly independent over $\mathds{Q}$ so that no two poles coincide, these poles are simple.

The double Gamma function is generalization of the Barnes G-function considered above and satisfies
\be 
\Gamma_2(z\, | \, 1,1)=\frac{(2\pi)^{\frac{z}{2}}}{G(z)}\ .
\ee
\subsection[The function \texorpdfstring{$\G_k(z)$}{Gk(z)}]{The function $\boldsymbol{\G_k(z)}$}
Finally, we define a function $\G_k(z)$ out of the double Gamma function that enters the unflowed three-point functions. It is given by\footnote{There is a typo in \cite{Maldacena:2001km} in the corresponding formula which we corrected here.}
\be 
\G_k(j)=(k-2)^{\frac{j(k-1+j)}{(k-2)}} \Gamma_2(-j \, |\, 1, k-2)\Gamma_2(k-1+j \, |\, 1, k-2)\ .
\ee
This is the main function that enters the three-point functions of the model. This function is sometimes confusingly also referred to as the Barnes double Gamma function. The following properties follow immediately from the definition and the functional equations of the double Gamma function:
\begin{subequations}
\begin{align}
\G_k(j+1)&=\lgamma\left(-\frac{j+1}{k-2}\right)\G_k(j)\ , \\
\G_k(j-k+2)&=\frac{\lgamma(j+1)}{(k-2)^{2j+1}} \G_k(j)\ , \\
\G_k(j)&=\G_k(-j+1-k)\ ,
\end{align} \label{eq:Barnes G double Gamma functional identities}
\end{subequations}
where we recall that $\lgamma(x)=\frac{\Gamma(x)}{\Gamma(1-x)}$. The function $\G_k(j)$ has simple poles (unless $k \in \mathds{Q}$ in which case some of the following poles will coincide) at
\be 
j=m+n(k-2)\ , \qquad j=-(m+1)-(n+1)(k-2)\ ,
\ee
for $m,n \in \mathds{Z}_{\ge 0}$.

We also notice that for $k=3$, this function is directly related to the usual Barnes G function as follows:
\be 
\G_3(z)=\Gamma_2(-z\, | \, 1,1)\Gamma_2(2+z\, | \, 1,1)=\frac{2\pi}{G(-z) G(z+2)}\ .
\ee

\section{Unflowed three-point functions} \label{app:unflowed three-point functions}

In this appendix, we review the form of the unflowed three-point functions as found in \cite{Teschner:1997ft, Teschner:1999ug}. We follow the conventions of \cite{Maldacena:2001km}. We write
\be 
D(j_1,j_2,j_3)=\left \langle V^0_{j_1}(0;0)V^0_{j_2}(1;1)V^0_{j_3}(\infty;\infty) \right \rangle\ .
\ee
We normalize vertex operators as
\begin{multline}
\left \langle V^0_{j_1}(x_1;z_1) V^0_{j_2}(x_2;z_2) \right \rangle\\
=\frac{1}{|z_1-z_2|^{4 \Delta_1}} \left(\delta^2(x_1-x_2) \delta(j_1+j_2-1)+\frac{B(j_1)}{|x_1-x_2|^{4j_1}} \delta(j_1-j_2) \right)\ . \label{eq:normalization vertex operators}
\end{multline}
$B(j)$ is then given by
\be 
B(j)=\frac{k-2}{\pi} \frac{\nu^{1-2j}}{\lgamma\left(\frac{2j-1}{k-2}\right)}\ . \label{eq:B}
\ee
$B(j)$ is related to $R_{j}$ as follows (by definition from \eqref{eq:Teschner-reflection} and \eqref{eq:normalization vertex operators}):
\be 
R_j=\frac{\pi B(j)}{2j-1}\ .\label{eq:reflection coefficient normalization}
\ee
The three-point function in $x$-space is given by
\be 
D(j_1,j_2,j_3)=-\frac{\G_k(1-j_1-j_2-j_3)\G_k(j_3-j_1-j_2)\G_k(j_2-j_1-j_3)\G_k(j_1-j_2-j_3)}{2\pi^2 \nu^{j_1+j_2+j_3-1} \lgamma\left(\frac{k-1}{k-2}\right) \G_k(-1) \G_k(1-2j_1)\G_k(1-2j_2)\G_k(1-2j_3)}\ . \label{eq:unflowed three point function}
\ee
Recall from the discussion in Section~\ref{sec:y-transform} that this already incorporates any combination of continuous and discrete representations.\footnote{This formula was derived in the $H_3^+$ model, which does not possess discrete representations. As argued in \cite{Maldacena:2001km}, three-point functions for discrete representations can be obtained by analytic continuation in the spins $j_i$. We will tacitly assume this assumption to hold.}

\subsection{Limits and identities}
One can recover the two-point function from the three-point function by considering the limit $j_3 \to 0$ as explained in \cite{Teschner:1999ug}. The result is that
\be 
D(j_1,j_2,0) =B(j_1) \delta(j_1-j_2)\ .
\ee
Similarly, one can show that 
\be  
D(\tfrac{k}{2}-j_1,j_2,\tfrac{k}{2}) =B(\tfrac{k}{2}) \, \delta(j_1-j_2)\ .
\ee
Making use of \eqref{eq:Barnes G double Gamma functional identities}, one can derive  the following useful identities
\begin{subequations}
\begin{align}
\mathcal N(j_1) D(\tfrac{k}{2}-j_1,j_2,j_3)&=\mathcal N(j_2) D(j_1,\tfrac{k}{2}-j_2,j_3)\nonumber\\
&=\mathcal N(j_3) D(j_1,j_2,\tfrac{k}{2}-j_3) \nonumber\\
&=\mathcal{N}(j_1)\mathcal{N}(j_2)\mathcal{N}(j_3) D(\tfrac{k}{2}-j_1,\tfrac{k}{2}-j_2,\tfrac{k}{2}-j_3)\ , \label{eq:symmetry shifted three point function}\\
D(j_1,j_2,j_3) &= \mathcal N(j_1) \mathcal N(j_2)  D(\tfrac{k}{2}-j_1,\tfrac{k}{2}-j_2,j_3)\nonumber \\ 
& =\mathcal N(j_2)\mathcal N(j_3) D(j_1,\tfrac{k}{2}-j_2,\tfrac{k}{2}-j_3) \nonumber \\
& =\mathcal N(j_1)\mathcal N(j_3) D(\tfrac{k}{2}-j_1,j_2,\tfrac{k}{2}-j_3)  \label{eq:symmetry shifted three point function 2}\\
D(1-j_1,j_2,j_3)&=B(1-j_1) \frac{\pi \, \lgamma(2j_1-1)\lgamma(-j_1+j_2-j_3+1)}{\lgamma(j_1+j_2-j_3)} D(j_1,j_2,j_3)\ .\label{eq:reflection symmetry unflowed three point function}
\end{align}
\end{subequations}

\section{Surface integral identities}
\label{app:integral-identities}

In this appendix, we discuss, various surface integral identities that we frequently use. We adopt the following convention for the integral measure, 
\be 
\text d^2 y \equiv \text{d}\Re(y) \,  \text{d} \Im(y) \ . 
\ee

\paragraph{Two singular points.} The basic identity for $a-\bar{a} \in \mathds{Z}$ is
\be 
\int \mathrm{d}^2 y \ y^{a-1} \bar{y}^{\bar{a}-1}=\pm i (2\pi)^2 \delta_{a,\bar{a}}\, \delta(a+\bar{a})\ ,
\label{eq:two-singular-points}
\ee
where $\delta_{a,\bar{a}}$ is the discrete Kronecker delta and $\delta(a+\bar{a})$ is the Dirac delta. We will frequently abbreviate $\delta_{a,\bar{a}} \, \delta(a+\bar{a})$ with the more compact notation $\delta^{(2)}(a)$. The sign is ambiguous. To prove this identity, one has to use analytic continuation and similarly in all the subsequent identities. We illustrate this procedure in this simple case. First, change to radial coordinates
\be 
\int \mathrm{d}^2 y \,  y^{a-1} \bar{y}^{\bar{a}-1}=2\pi \delta_{a,\bar{a}} \int_0^\infty \mathrm{d}r \ r^{a+\bar{a}-1}\ .
\ee
The latter integral is obviously never convergent and we use analytic continuation to give it a value. Changing coordinates to $x=\log(r)$, we need to define the integral
\be 
\int \mathrm{d} x\ \mathrm{e}^{x(a+\bar{a})}\ .
\ee
We can rotate the contour in $x$ to obtain the analytically continued form
\be 
\pm i \int \mathrm{d} x\ \mathrm{e}^{i x(a+\bar{a})}=\pm 2\pi i \delta(a+\bar{a})\ .
\ee
The sign depends on the direction in which we rotate the contour. 

Since the integral is subtle, we also mention a different way to obtain the same result. Introduce a regulator in the integral as follows:
\be 
\int \mathrm{d} x\ \mathrm{e}^{x(a+\bar{a})-\varepsilon x^2}=\sqrt{\frac{\pi}{\varepsilon}} \ \mathrm{e}^{\frac{(a+\bar{a})^2}{4\varepsilon}}\ .
\ee
The integral converges for $\Re (\varepsilon)>0$. We then analytically continue the result in $\varepsilon$. The resulting Gaussian provides an approximation of the Dirac delta for $\Re( \varepsilon)<0$. Taking the square root is ambiguous and we obtain again the result
\be 
\int \mathrm{d} x\ \mathrm{e}^{x(a+\bar{a})}=\pm 2\pi i \delta(a+\bar{a})\ ,
\ee
corresponding to the two different analytic continuations.
\paragraph{Three singular points.} Another surface integral that we often encounter is
\be 
\int \mathrm{d}^2 y \ y^{a-1} \bar{y}^{\bar{a}-1} (1-y)^{b-1}(1-\bar{y})^{\bar{b}-1}=\frac{\pi \lgamma(a)\lgamma(b)}{\lgamma(a+b)}\ ,
\label{eq:three-singular-points}
\ee
where
\be 
\lgamma(x)=\frac{\Gamma(x)}{\Gamma(1-\bar{x})}\ .
\ee
By $\bar{x}$ we mean the corresponding right-moving quantity, which is not the complex conjugate of $x$. 
The derivation of this formula is classic in the context of the Coulomb gas formalism \cite{Dotsenko:1984nm} or the KLT formula \cite{Kawai:1985xq}.

We also need the following higher-dimensional generalization of this result:
\be 
\int \prod_{i=1}^n \mathrm{d}^2 y_i \ y_i^{a_i-1} \bar{y}_i^{\bar{a}_i-1} \left(1-\sum_{j=1}^n y_j \right)^{b-1}\left(1-\sum_{j=1}^n \bar{y}_j \right)^{\bar{b}-1}=\frac{\pi^n \lgamma(b)\prod_{i=1}^n \lgamma(a_i)}{\lgamma\left(b+\sum_{i=1}^n a_i\right)}\ ,
\label{general-integral}
\ee
which can be proved recursively by the substitution $y_n=x_n\left(1-\sum_{i=1}^{n-1} y_i\right)$ of $y_n$.

\section{Explicit form of the three-point functions of three continuous representations}
\label{app:integral Lauricella}

In this appendix, we explicitly perform the integral over $y$-space in order to obtain the form of the three-point function in physical $h$-space for three continuous representations. We are able to do this computation only in the special case where
\be 
w_i \ge 1\ , \qquad w_1+w_2+w_3\in 2\mathds{Z}+1\ , \qquad w_1+w_2+w_3 \ge 2w_i+1 \label{eq:spectral flow indices range Lauricella}
\ee
for all $i$. The restriction $w_i \ge 1$ is only imposed to avoid case distinctions, since we would need to drop integrations over the unflowed vertex operators in the following. We have seen that the three-point function is given by the following integral \eqref{eq:three point function integral expression},
\begin{multline}
\hspace{-10pt}\left\langle \prod_{i=1}^3 V_{j_i,h_i,\bar{h}_i}^{w_i}(x_i;z_i) \right\rangle= \int \prod_{i=1}^3 \mathrm{d}^2 t_i \ \left|t_i^{-\frac{k w_i}{2}+j_i+h_i-1} (1-t_i)^{\frac{k w_i}{2}+j_i-h_i-1}\right|^2\\
\times\left|1-\frac{w_1 t_1}{N}-\frac{w_2 t_2}{N}-\frac{w_3 t_3}{N}\right|^{2(\frac{k}{2}-j_1-j_2-j_3)}\ ,
\end{multline}
where $N=\frac{1}{2}(w_1+w_2+w_3-1)$ and up to a prefactor that we will reinstate at the end. Here we have already performed the change of variables that is discussed in Section~\ref{subsec:three point function h dependence}. Here and in the following we use the short hand $|z^a|^2=z^a\bar{z}^{\bar{a}}$. We do not assume that $a=\bar{a}$, but rather $a-\bar{a} \in \mathds{Z}$. This integral is of hypergeometric type. Our strategy will be to make use of twisted (co)homology, which we very briefly review in the following. For more extensive references, see \cite{Aomoto}. For applications to string theory, see \cite{Mizera:2019gea} and references therein.
\subsection{Twisted (co)homology}
Let $u$ be a multivalued function on $\mathds{C}^m$ with singularities on a codimension 1 sublocus $S$. Let $M=\mathds{C}^m \setminus S$. We are interested in computing integrals of the form
\be 
\int_\Delta \varphi u\ ,
\ee
where $\varphi$ is a single-valued $m$-form and $\Delta$ a cycle. Since $u$ is multivalued, we also have to specify the branch of $u$ we want to use. This is done by \emph{loading} $\Delta$, resulting in a twisted cocycle of the form $\Delta \otimes u_\Delta$, where $u_\Delta$ labels the branch. 

One can develop a (co)homology theory for the twisted cocycle $\varphi$ and the twisted cycle $\Delta$. Without going into details, we list here the most important definitions and properties.
\paragraph{Definition of the cohomology groups.} The space of cycles $\varphi$ is equipped with a twisted differential 
\be 
\nabla_\omega\varphi=(\mathrm{d}+\omega \wedge)\varphi\ ,
\ee
where $\omega=\mathrm{d} \log (u)$.  $\nabla_\omega$ squares to zero and hence defines twisted deRham groups:
\be 
H^k(M,\nabla_\omega)=\frac{\ker \nabla_\omega}{\mathop{\text{im}} \nabla_\omega}\ .
\ee
\paragraph{Definition of the homology groups.} Similarly, there is a dual notion of twisted homology groups. For a twisted cycle $\Delta \otimes u_\Delta$ one can define a boundary operator $\partial_\omega$, since a given branch of $u$ on $\Delta$ naturally restricts to a branch on its boundary. Since we also have $\partial_\omega^2=0$, this defines homology groups
\be 
H_k(M,\mathcal{L}_\omega)=\frac{\ker \partial_\omega}{\mathop{\text{im}} \partial_\omega}\ .
\ee
  \paragraph{Vanishing of homology groups.} Contrary to ordinary deRham (co)homology, most homology groups vanish. Generically, only the top homology group is non-vanishing, i.e. $H_k(M,\mathcal{L}_\omega)=\{0\}$ for $k<m$.
  \paragraph{Poincar\'e duality.} There is a notion of Poincar\'e duality for twisted (co)homology. There is a natural pairing
  \begin{subequations}
  \begin{align}
  H_m(M,\mathcal{L}_\omega) \times H^m (M,\nabla_\omega) &\longrightarrow \mathds{C}\ , \\
  \langle \Delta \otimes u_\Delta, \varphi \rangle&=\int_{\Delta\otimes u_\Delta} \varphi(z) \equiv \int_\Delta (\text{branch $u_\Delta$ of $u$}) \varphi\ .
  \end{align}
  \end{subequations}
  It is known that this pairing is non-degenerate and consequently there is an isomorphism $H_m(M,\mathcal{L}_\omega)\allowbreak \cong \allowbreak H^m(M,\nabla_\omega)$.
  \paragraph{Compactly supported cohomology.} The manifold $M$ is non-compact, so in general the above pairing might not be well-defined. For the pairing to be well-defined, one needs that either $\Delta$ is compact or $\varphi$ is compactly supported. There is a map
  \be 
  \iota_\omega: H^m(M,\nabla_\omega) \longrightarrow H^m_c(M,\nabla_\omega)
  \ee
  from cohomology to compactly supported cohomology defined by picking another cocycle that is compactly supported, but cohomologous to the original one. 
  \paragraph{Dual bundle.} We can also define (co)homology groups obtained by replacing $\omega$ with $\bar{\omega}$. We call the corresponding bundle $\mathcal{L}^\vee_\omega=\mathcal{L}_{\bar{\omega}}$. These are the corresponding dual (co)homology groups. There exists another notion of dual bundle that is employed more often in the literature and is obtained by replacing $\omega \to -\omega$. The notions are different, but many properties carry over.
  \paragraph{Intersection forms.} There are intersection pairings between the groups $H_m(M,\mathcal{L}_\omega)$ and $H_m(M,\mathcal{L}_\omega^\vee)$ and between $H^m(M,\nabla_\omega)$ and $H^m(M,\nabla_{\bar{\omega}})$. The latter is given by
  \be 
  \langle \varphi_i(z)|\varphi_j^\vee(z) \rangle=\int_M |u(z)|^2 \varphi_i(z) \wedge \overline{\varphi_j^\vee(z)}\ ,
  \ee
  provided that the integral converges. The intersection number between homologies is defined as
  \be 
  \langle \Delta \otimes u_\Delta | \Delta^\vee \otimes \overline{u_{\Delta^\vee}}\rangle=\sum_{z \in \Delta\cap \Delta^\vee} \langle \Delta | \Delta^\vee \rangle_z \frac{u_\Delta(z)\overline{u_{\Delta^\vee}(z)}}{|u(z)|^2}=\sum_{z \in \Delta\cap \Delta^\vee} \langle \Delta | \Delta^\vee \rangle_z u_\Delta(z)u_{\Delta^\vee}(z)^{-1}\ ,
  \ee
  where $\langle \Delta | \Delta^\vee \rangle_z$ is the topological intersection number at $z$.
  \paragraph{Homological splitting.} Let $\Delta_\alpha \otimes u_{\Delta_\alpha}$ be a basis of $H_m(M,\mathcal{L}_\omega)$. We collect the intersection products in the intersection matrix:
  \be 
  \mathbf{I}_{\alpha\beta}=\langle \Delta_\alpha \otimes u_{\Delta_\alpha} | \Delta_\beta^\vee \otimes u_{\Delta_\beta^\vee}\rangle\ .
  \ee
  For the purpose of intersection theory, the following combination is the identity:
  \be 
   \sum_{\alpha,\beta} (\mathbf{I}^{-1})_{\beta\alpha} | \Delta_\beta^\vee \otimes u_{\Delta_\beta^\vee}\rangle \langle \Delta_\alpha \otimes u_{\Delta_\alpha} | \ .
  \ee
  Thus we can insert it in the cohomological intersection pairing to homologically split the integral:
  \be 
  \langle \varphi | \varphi^\vee \rangle =   \sum_{\alpha,\beta} (\mathbf{I}^{-1})_{\beta\alpha} \langle \varphi^\vee , \Delta_\beta^\vee \otimes u_{\Delta_\beta^\vee}\rangle \langle \Delta_\alpha \otimes u_{\Delta_\alpha} , \varphi \rangle\ .
  \ee
  This reduces the integral to contour integrals.
  
\subsection[The case of the Lauricella hypergeometric function \texorpdfstring{$F_A$}{FA}]{The case of the Lauricella hypergeometric function $\boldsymbol{F_A}$}
  We specialize the previous discussion to the case at hand:
  \begin{align} 
  u&=\prod_k t_k^{b_k} (1-t_i)^{c_k-b_k-1} \left(1-\sum_k x_k t_k \right)^{-a}\ , \\
  S&= \left \{ \prod_{k=1}^m x_k  \prod_{\{i_1,\dots,i_r\}\subset \{1,\dots,m\}} \left(1-\sum_{p=1}^r x_{i_p} \right)=0 \right\} 
  \end{align}
  and $M=\mathds{C}^m \setminus S$. We ultimately want to set $m=3$ and specify $a$, $b_k$ and $c_k$, but we keep them general for the moment. This system was analyzed by Goto \cite{Goto} and in the following, we heavily rely on the results that were obtained there.
  We have $\dim H_m(M,\mathcal{L}_\omega)=2^m$. A basis of this homology group is given by the twisted cycles $\Delta_{i_1 \dots i_r}$, where $\{i_1,\dots,i_r\} \subset \{1,\dots,m\}$ runs over all subsets. We refer to \cite{Goto} for the precise definition of $\Delta_{i_1\cdots i_r}$, since we will not need it in the following.
  
  We are interested in computing the intersection pairing $\langle \varphi | \varphi \rangle$
  where
  \be 
  \varphi=\frac{\mathrm{d} t_1 \wedge \cdots \wedge \mathrm{d} t_m}{t_1 \cdots t_m}\ .
  \ee
  The intersection numbers of $\Delta_{i_1,\dots,i_r}$ were computed in \cite{Goto}. We have $\langle \Delta_I | \Delta_J^\vee \rangle=0$ for $I \ne J$.\footnote{\cite{Goto} used a different definition of dual bundle. But the definition of intersection number agrees, since all the exponents of $u$ are real. } The self-intersection reads
  \begin{align} 
 \mathbf{I}_{i_1,\dots,i_r}&= \langle\Delta_{i_1,\dots,i_r}|\Delta^\vee_{i_1,\dots,i_r} \rangle=\frac{\alpha-\prod_p \gamma_{i_p}}{(\alpha-1)\prod_p (1-\gamma_{i_p})} \prod_q \frac{\beta_{j_q}(1-\gamma_{j_q})}{(1-\beta_{j_q})(\beta_{j_q}-\gamma_{j_q})}\\
 &=(-1)^m \frac{\sin\left(\pi\left(a-\sum_p c_{i_p}\right)\right)}{\sin(\pi a) \prod_p \sin (\pi c_{i_p})} \prod_q \frac{\sin(\pi c_{j_q})}{\sin(\pi b_{j_q}) \sin (\pi (c_{q_j}-b_{q_j}))}\ ,
  \end{align}
  where $\alpha=\mathrm{e}^{2\pi i a}$, $\beta_i=\mathrm{e}^{2\pi i b_i}$, $\gamma_i=\mathrm{e}^{2\pi i c_i}$. Products over $p$ run from $1,\dots,r$, so that $i_p$ runs over the subset $I=\{i_1,\dots,i_r\}$. Products over $q$ run from $1,\dots,m-r$ so that $j_q$ runs over the complement $\{1,\dots,m\} \setminus I$. The homological splitting hence reads
  \be 
  \langle \varphi | \varphi \rangle=\sum_{\{i_1,\dots,i_r\} \subset \{1,\dots,m\}}  \mathbf{I}_{i_1,\dots,i_r}^{-1} \left|\int_{\Delta_{i_1,\dots,i_r}} u \varphi\right|^2
  \ee
  The latter integral is also evaluated in \cite{Goto} and gives
  \begin{multline} 
  \int_{\Delta_{i_1,\dots,i_r}} u\varphi=\mathrm{e}^{\pi i \left(\sum_p b_{i_p}-\sum_p c_{i_p}+r\right)} \prod_{p} x_{i_p}^{1-c_{i_p}} \prod_{p} \Gamma(c_{i_p}-1) \prod_q \frac{\Gamma(b_{j_q}) \Gamma(c_{j_q}-b_{j_q})}{\Gamma(c_{j_q})} \\
  \times\frac{\Gamma(1-a)}{\Gamma(\sum_p c_{i_p}-a-r+1)} F_A(a+r-\sum_{p=1}^r c_{i_p},b^{i_1\cdots i_r},c^{i_1\cdots i_r};x)\ ,
  \end{multline}
  where
  \be 
  b^{i_1\cdots i_r}=b+\sum_p (1-c_{i_p})e_{i_p}\ , \qquad  c^{i_1\cdots i_r}=c+2\sum_p (1-c_{i_p})e_{i_p}
  \ee
  and $e_i$ is the unit vector in the $m$-th direction. $F_A$ is the Lauricella hypergeometric function of type A defined in eq.~\eqref{eq:definition Lauricella hypergeometric function}. Putting these pieces together, we hence obtain
  \begin{align}
  &\int \prod_{k=1}^m \mathrm{d}^2 t_k \ \left|\prod_k t_k^{b_k-1} (1-t_i)^{c_k-b_k-1}  \left(1-\sum_k x_k t_k \right)^{-a}\right|^2\nonumber\\
 &\qquad =\sum_{\{i_1,\dots,i_r\} \subset \{1,\dots,m\}} \frac{\sin(\pi a) \prod_p \sin (\pi c_{i_p})}{\sin\left(\pi\left(a-\sum_p c_{i_p}\right)\right)} \prod_q \frac{\sin(\pi b_{j_q}) \sin (\pi (c_{q_j}-b_{q_j}))}{\sin(\pi c_{j_q})} \nonumber\\
 &\qquad\qquad\times\Bigg|\mathrm{e}^{\pi i \left(\sum_p b_{i_p}-\sum_p c_{i_p}\right)} \prod_{p} x_{i_p}^{1-c_{i_p}} \prod_{p} \Gamma(c_{i_p}-1) \prod_q \frac{\Gamma(b_{j_q}) \Gamma(c_{j_q}-b_{j_q})}{\Gamma(c_{j_q})} \nonumber\\
& \qquad\qquad \times\frac{\Gamma(1-a)}{\Gamma(\sum_p c_{i_p}-a-r+1)}  F_A(a+r-\sum_{p=1}^r c_{i_p},b^{i_1\cdots i_r},c^{i_1\cdots i_r};x)\Bigg|^2\ .
  \end{align}
By construction the intersection kernel does not depend on whether we choose $b$ or $\bar{b}$, since it is periodic in $b \to b+1$ and similarly for the other variables.
  
  We can then rewrite the $\Gamma$-functions of the right-movers by using the reflection formula of the $\Gamma$-function. This leads to
    \begin{align}
  &\int \prod_{k=1}^m \mathrm{d}^2 t_k \ \left|\prod_k t_k^{b_k-1} (1-t_i)^{c_k-b_k-1}  \left(1-\sum_k x_k t_k \right)^{-a}\right|^2\nonumber\\
 &\qquad = \pi^m \sum_{\{i_1,\dots,i_r\} \subset \{1,\dots,m\}}\frac{\lgamma(1-a)}{\lgamma(\sum_p c_{i_p}-a-r+1)} \prod_{p} \lgamma(c_{i_p}-1) \prod_q \frac{\lgamma(b_{j_q}) }{\lgamma(c_{j_q})\lgamma(1-c_{j_q}+b_{j_q})} \nonumber \\
&\qquad\qquad \times \Bigg|\prod_{p} x_{i_p}^{1-c_{i_p}}  F_A\left(a+r-\sum_{p=1}^r c_{i_p},b^{i_1\cdots i_r},c^{i_1\cdots i_r};x\right)\Bigg|^2\ ,
  \end{align}
where the definition of the $\lgamma$-function is given in eq.~\eqref{eq:gamma function}. This formula is symmetric in left- and right-movers despite appearances. It also generalizes the formulas given in Appendix~\ref{app:integral-identities} and reduces to them once one specifies the parameters accordingly.

Let us also note as a consistency check that in the case $m=1$ this formula is well-known and routinely used in the Coulomb gas formalism \cite{Dotsenko:1984nm}. It reduces to
  \begin{multline}
    \int \mathrm{d}^2 t \ \left|t^{b-1} (1-t)^{c-b-1}  (1- x t )^{-a}\right|^2= \frac{\pi \lgamma(b)\lgamma(c-b)}{\lgamma(c)}\left| {}_2F_1 \left(a,b,c;x\right)\right|^2\\
    +\frac{\pi  \lgamma(c-1)\lgamma(1-a)}{\lgamma(c-a)} \left|x^{1-c} {}_2F_1 \left(a-c+1,b+1-c,2-c;x\right)\right|^2\ ,
  \end{multline}
  which matches e.g.~with the formulas in \cite{DiFrancesco:1997nk}. 
  \subsection{Full three-point function}
  Finally, we use this formula to derive the three point function. We specify $x=\frac{\boldsymbol{w}}{N}$ (where $\boldsymbol{w}$ is viewed as a vector) and
  \begin{subequations}
  \begin{align}
  b_i&=-\frac{k w_i}{2}+j_i+h_i\ , \\
  c_i&=2j_i \ , \\
  a&=-\frac{k}{2}+j_1+j_2+j_3\ .
  \end{align}
  \end{subequations}
  For this choice, we have
  \begin{align}
  \frac{\lgamma(b_i)}{\lgamma(c_i)\lgamma(1-c_i+b_i)}=\frac{\lgamma(h_i+j_i-\frac{kw_i}{2})}{\lgamma(2j_i)\lgamma(1-j_i+h_i-\frac{kw_i}{2})}=\frac{R(j_i,h_i,\bar{h}_i)}{R_{j_i}(2j_i-1)}
  \end{align}
  We obtain the following three-point function
    \begin{align}
 &\left\langle \prod_{i=1}^3 V_{j_i,h_i,\bar{h}_i}^{w_i}(x_i;z_i) \right\rangle= \pi^3 C(w_1,w_2,w_3;j_1,j_2,j_3;k)  \prod_{i=1}^3 a_i^{\frac{k}{2}(w_i-1)-h_i-\bar{h}_i} \Pi^{-k}  \nonumber\\
 &\qquad\times\lgamma\left(\frac{k}{2}-j_1-j_2-j_3+1\right)\sum_{\{i_1,\dots,i_r\} \subset\{1,2,3\}} \prod_p \lgamma(2j_{i_p}-1) \prod_q \frac{R(j_{j_q},h_{j_q},\bar{h}_{j_q})}{(2j_{j_q}-1)R_{j_{j_q}}} \nonumber\\
&\qquad\times N^{k-2j_1-2j_2-2j_3}\prod_{i=1}^3 w_i^{-\frac{k}{2}(w_i+1)+2j_i} \nonumber\\
&\qquad\times\frac{\left| F_A \left(-\frac{k}{2}+j_1+j_2+j_3,-\frac{k w_i}{2}+j_i+h_i,2j_i;\frac{w_i}{N} \right)\right|^2}{\lgamma(\frac{k}{2}-j_1-j_2-j_3+1)}\Bigg|_{j_{i_r} \to 1-j_{i_r}}\ .
\label{h-basis-3pt-C}
  \end{align}
  The replacement only applies to the last two lines of this formula.  The definitions of $a_i$ and $\Pi$ were given in \eqref{eq:three-pt-ai} and \eqref{eq:definition C}. 
The conjecture in Section~\ref{sec:normalisation-conjecture} means that the constant prefactor $C(w_1,w_2,w_3;j_1,j_2,j_3;k)$ is in fact given by the unflowed three-point function $\mathcal N(j_1)D(\tfrac{k}{2}-j_1,j_2,j_3)$ of eq.~\eqref{eq:unflowed three point function}. As a consistency check of our computation one can check that this result indeed reflection symmetric when making use of \eqref{eq:reflection symmetry unflowed three point function}. This determines the spectrally flowed three-point functions completely for the range \eqref{eq:spectral flow indices range Lauricella}. 

\bibliographystyle{JHEP}
\bibliography{bib}
\end{document}